\documentclass[a4paper,11pt]{article}
\pdfoutput=1 

\usepackage{jcappub} 
\usepackage{comment}
\usepackage{orcidlink}

\usepackage[T1]{fontenc} 
\usepackage{float}
\usepackage{subcaption}
\usepackage{caption} 
\usepackage{natbib,xcolor,amsmath,amssymb,bm,multirow}

\newcommand{\orcid}[1]{\orcidlink{#1}}

\title{\boldmath One Halo, Two Boundaries: Relating Accretion Shocks and Splashback Radii in Galaxy Clusters}

\author[1,a]{Siddhant Sen,\note{Corresponding author.}}
\author[a]{Susmita Adhikari,\orcid{0000-0002-0298-4432}}
\author[b]{Daisuke Nagai,\orcid{0000-0002-6766-5942}}
\author[c]{Benedikt Diemer \orcid{0000-0001-9568-7287}}

\affiliation[a]{Department of Physics, Indian Institute of Science Education and Research Pune, Dr. Homi Bhabha Road, Pashan, Pune, India}
\affiliation[b]{Department of Physics, Yale University, New Haven, USA}
\affiliation[c]{Department of Astronomy, University of Maryland, College Park MD, USA}

\emailAdd{siddhant.sen@students.iiserpune.ac.in}
\emailAdd{susmita@iiserpune.ac.in}
\emailAdd{daisuke.nagai@yale.edu}
\emailAdd{diemer@umd.edu}
\abstract{The boundaries of dark matter and gas in clusters are delineated by the splashback radius and the accretion shock, respectively. Theoretically, both of these boundaries are expected to coincide at the outskirts of halos. However, hydrodynamic cosmological simulations have highlighted significant displacement between them. In this study, we utilize the IllustrisTNG simulation suite to investigate the statistical relationship between the splashback and shock surfaces in a sample of 812 cluster-mass halos. We compute the full angular distribution of both boundaries and examine their relationship, also considering how different moments of this distribution correlate with halo properties. We employ a dispersion-based measure for the splashback boundary and the maximum entropy distance for the shock location. Despite examining various boundary definitions, we consistently observe an offset between the splashback and shock boundaries, with $R_{\rm sh}/R_{\rm sp} \sim 1.3-2$, depending on specific methodological choices. This offset predominantly occurs along void directions. We analyze the redshift evolution of these boundaries for a subset of halos and find that splashback and shock boundaries are not necessarily distinct at earlier times. During mergers, gas dissipates energy and resists contraction via pressure, unlike collisionless dark matter, leading to the observed boundary offset. We also find that the feature in pressure profiles arising from the outer accretion shock is sensitive to the exact method of stacking, which has important implications for observations.}

\begin{document}
\maketitle
\section{Introduction}
\label{sec:intro}

The boundaries of dark matter halos are usually defined at a radius that encloses a reference overdensity compared to the background universe. However, it has now been established that a more dynamically motivated boundary can be defined for dark matter particles based on the splashback radius, which is the first apocenter of the most recently accreted model \cite{Diemer:2014xya, Adhikari:2014lna, More:2015ufa}. Since this is an apocenter and particles are turning around at this location, this also corresponds to the outermost density caustic in the dark matter halo profile and separates the infall and multi-streaming regions \cite{FG84,Bertschinger:1985pd,Diemer:2014xya, Adhikari:2014lna}.

Unlike collisionless dark matter particles, the infalling gas does not orbit and is instead expected to stall during infall due to its encounter with the hot intracluster gas; This causes the gas to shock and convert its kinetic energy into heat, marked by a strong jump in temperature, pressure, and entropy \citep{Bertschinger:1985pd, 2003ApJ...593..599R,Skillman_2008,Molnar_2009}. In analytical self-similar models of the secondary collapse of gas and dark matter, such as those explored in \citep{Bertschinger:1985pd, GunnGott72, Shi:2016xeu}, it has been shown that the outer caustic in dark matter coincides with the shock radius for gas that follows an adiabatic equation of state \citep{Shi:2016xeu}. This shock front, often referred to as the accretion shock, is thought to define the outer boundary of the hot gaseous halo. Both the shock and splashback radii are expected to probe the mass and accretion history of a growing peak \citep{Diemer:2014xya, Adhikari:2014lna,Chue:2018hxk, Shin:2022iza, Lucie-Smith:2022mar,Lucie-Smith:2023kue, Shi:2016lwp,Velmani:2023hxk, R:2024idl,Dacunha:2025elg}. Specifically, more massive halos are expected to have larger boundaries; at fixed mass, a higher accretion rate pushes the boundary to smaller radii in units of $R_{\rm 200m}$ (i.e., corresponding to higher overdensities). Additionally, it has been shown that the locations of these boundaries encode important information about the nature of gravity, dark matter, and cosmological parameters \citep{Adhikari:2018izo, Contigiani:2018hbn, Banerjee:2019bjp, Haggar:2024rft}.

While simple spherical self-similar theoretical models predict that these boundaries should coincide, recent work has shown that the splashback radius for dark matter and the shock radius for gas can be significantly offset from each other in cosmological hydrodynamical simulations \cite{Aung:2020grs, Walker:2018fqw}. In the \textit{Omega500} simulations of 60 massive clusters, the shock radius was found to be, on average, as much as $\sim 2$ times larger than the splashback radius \cite{Aung:2020grs}. A similar offset was also found in \cite{Zhang:2024tue}, who studied the relationship between shock and splashback in the ThreeHundred simulations \citep{2018MNRAS.480.2898C}. The stacked thermal Sunyaev-Zeldovich (SZ) effect \citep{1972CoASP...4..173S} profiles of halos in the ThreeHundred simulations were studied to detect shocks in SZ surveys \citep{Baxter:2021tjr}. 

A possible explanation for the offset was provided by \cite{Zhang:2019kej}, who showed that such an offset can be expected due to runaway merger shocks that lead to the acceleration of the accretion shock when they encounter each other. While mergers are ubiquitous in structure formation, given the shape of the halo mass function, large mergers are rare \citep{1991ApJ...379..440B, 1994MNRAS.271..676L,2007IJMPD..16..763Z,2022ApJ...929..120D, 2025ApJ...988..160J}. A study of these effects using the \textit{Omega500} simulations found that the connection between mergers and shock expansion remains unclear, as the offsets appear to have been set significantly early, around z $\approx$ 4 during the formation of the halo \cite{Aung:2020grs}. 

Beyond its theoretical interest, this offset has important practical implications. The magnitude of this offset functions as a cosmological probe, retaining an imprint of the halo's complete merger history. Observationally, interpreting gas-based boundaries (SZ/X-ray) as direct tracers of the dark matter splashback radius leads to systematic biases in estimates of halo masses, concentration parameters, etc. The offset we measure provides a simulation-calibrated correction for observational studies. This discrepancy motivates a comprehensive investigation using a large, statistically representative cluster sample with modern galaxy formation physics.

In this work, we comprehensively explore the connection between the shock radius and splashback radius in the IllustrisTNG simulation suite, including both TNG300-1 and TNG--Cluster simulations \cite{Nelson-TNG-cl, Nelson:2018uso}. Our analysis covers the entire cluster mass range using a significantly larger sample: 460 galaxy clusters from a 300~Mpc uniform cosmological box and 352 zoom re-simulations of galaxy clusters of a cosmological population. The simulations use the same complete baryonic physics modeling as that of Illustris--TNG \cite{2018MNRAS.475..648P, 2018MNRAS.475..624N}. These increase the sample of halos compared to the \textit{Omega500} simulations by a factor of 13 and provide an avenue to carefully revisit the boundary relations using a large, representative cosmological sample in the moving-mesh AREPO code \cite{Weinberger:2019tbd}.

As halo boundaries are inherently aspherical \citep{2002ApJ...574..538J, 2006MNRAS.367.1781A}, we incorporate into our analysis the full directional distribution of the shock and splashback surfaces of each individual cluster. Although we often summarize the splashback radius or the shock radius as a single averaged quantity, the angular distribution of these boundaries carries important information about the environment and accretion history and can play a key role in refining our understanding of the subtleties of gas and DM physics in simulations. In addition, we study how the definition of the boundaries affects our inferences and how these quantities evolve with redshift.

The observational detection of halo boundaries is now a mature field, with the dark matter splashback feature definitively measured by multiple teams \cite{More:2016vgs,Baxter:2017csy,Shin19,Zuercher:2018prq,Murata:2020enz, DES:2021qzb, 2021ApJ...923...37A}. Attention is now shifting to the more elusive gas boundary, marked by the recent statistical detection of the accretion shock via the thermal Sunyaev-Zel'dovich (tSZ) effect \cite{Anbajagane:2021bnx}. Observing these boundaries in tandem unlocks the cluster outskirts as a laboratory where comparing their relative locations provides a pristine probe of differential gas/dark matter dynamics, while the state of the gas itself reveals the impact of microphysical processes like electron-proton equilibration \cite{Rudd:2009,Avestruz:2014dea}. A complete understanding of this `gastrophysics' is the prerequisite for unlocking the full cosmological potential of clusters, especially as powerful X-ray missions like SRG/eROSITA \cite{Zhang:2025eyu} and novel probes like the kSZ effect and FRBs are poised to deliver high-fidelity data. As these surveys begin to probe the gas boundary with increasing precision, a robust theoretical understanding of its relationship to the underlying dark matter halo, as explored in this paper, is imperative.

Our paper is organised as follows: in section 2, we describe the simulations used. In section 3, we describe the methods; in particular, the shock and splashback finding algorithm. in section 4, we discuss the results, and in section 5, we summarize our conclusions.
\section{Simulations}
Our analysis is built upon a synergistic framework that leverages the complementary strengths of the TNG-Cluster and TNG300 simulations. This dual approach allows us to combine the high-fidelity physics of massive clusters with a broad cosmological context and rigorous tests for systematic effects.

We analyze the cluster mass halos from the TNG-Cluster simulation \cite{Nelson-TNG-cl}, which is a suite of 352 multi-mass ``zoom'' re-simulations of cluster halos, simulated with gravo - magnetohydrodynamics (MHD) and incorporating a comprehensive model for galaxy formation physics\cite{Weinberger:2016ueb,Pillepich:2017jle}. In particular, TNG uses the AREPO code \cite{Weinberger:2019tbd}, which solves for the coupled evolution under self-gravity and ideal, continuum MHD \cite{Pakmor:2011ht,10.1093/mnras/stt428}. Self gravity in TNG is solved with a Tree-PM approach. TNG-Cluster also adopts the same TNG cosmology \cite{Planck:2015fie}: 
$\Omega_{m} = 0.3089, ~\Omega_{b} = 0.0486, ~\Omega_{\Lambda} = 0.6911,~ H_{0} = 67.74\,\rm{km\,s}^{-1}\,\rm{Mpc}^{-1},~\sigma_8 = 0.8159,~ n_s = 0.9667$. TNG-Cluster samples halos equally in the high mass bin, which differentiates it from the cosmological halo mass function. The resulting cluster sample is $(1~\rm{Gpc})$ volume complete at $M_{200c} \geq 10^{15}M_\odot$
and compensates for the rapid drop-off in statistics in TNG300 for $M_{200c} > 10^{14.5}M _\odot$. The TNG-Cluster simulation has the same resolution as TNG300-1, with $m_{\rm{gas}} = 1.2 \times 10^7\,M_\odot,\quad m_{\rm{DM}} = 6.1 \times 10^7\,M_\odot$, and a spatial resolution of $\sim 1$~kpc.

Conversely, to place our findings into a broader context and validate their robustness, we utilise the flagship TNG300 cosmological box. TNG300 is a $\sim300$ Mpc periodic volume that follows the same TNG physics and cosmology as TNG-Cluster. The simulation suite includes three resolution levels: TNG300-1 (highest resolution, equivalent to TNG-Cluster), TNG300-2 ($\sim8\times$ lower mass resolution), and TNG300-3 ($\sim64\times$ lower mass resolution). While TNG-Cluster provides our high-fidelity, high-mass sample, TNG300 serves three indispensable roles in this work: $(1)$ it allows us to test the mass dependence of our findings by providing a statistically complete sample extending down to the group scale; $(2)$ its volume-limited nature provides a crucial baseline for quantifying any selection effects in the curated TNG-Cluster sample; and $(3)$ its availability at multiple resolution levels enables direct numerical convergence tests, ensuring our results are not resolution-dependent.

Our fiducial halo sample therefore consists of $352$ halos from the TNG-Cluster simulation; these clusters have $\log {M_{\rm 200m}/M_\odot}> 14.4$ and $460$ halos above $\log M_{\rm 200m}/M_\odot > 14$ from the TNG300-1 suite. Halos are identified using the standard friends-of-friends $\text{(FoF)}$ algorithm with a linking length of $b = 0.2$. Substructures within $\text{(FoF)}$ halos are identified using the SUB-FIND algorithm, and we use the SUBLINK \cite{Rodriguez-Gomez:2015aua} merger trees to link them across time. 

We extract the dark matter and gas particles from the simulation snapshots over a range of redshifts within $20 \rm Mpc$ of the halo centre. In particular, for the gas, we use the entropy, temperature, and pressure information. In the next section, we discuss our shock and splashback finding algorithm.

\section{Methods}
\label{sec:shock_algorithm}

\subsection{Shock finding algorithm}

\begin{figure}[t] 
    \centering
    \includegraphics[width=0.98\columnwidth]{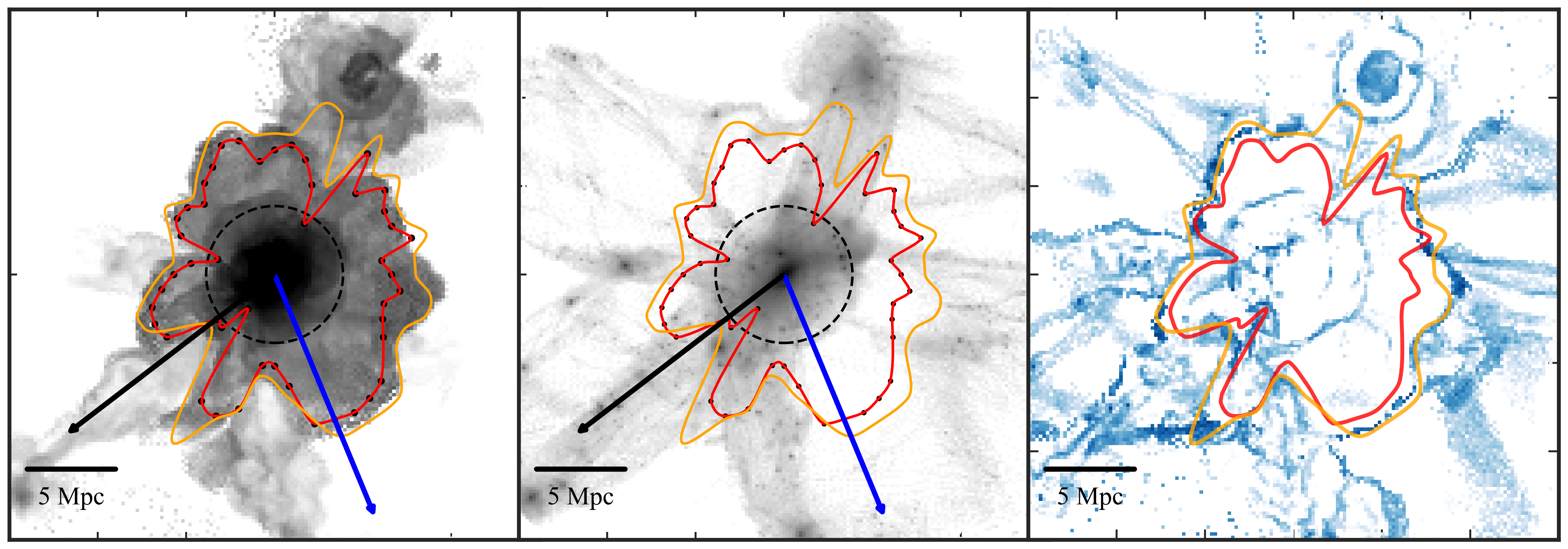}  
    \caption{A 2-dimension (2D) projection of the shock boundary found by our algorithm for one of the massive clusters from the TNG-Cluster suite with \(M_{200m}=1.9 \times 10^{15}\, M_\odot\).
    \emph{Left panel:} Shock boundaries found from entropy maxima (\emph{red}) and the logarithmic slope of entropy (\emph{orange}), overplotted on the gas entropy. A particular sightline along the void is shown with the blue solid arrow and along the filament by the black solid arrow.
    \emph{Middle panel:} Shock boundaries overplotted on DM density with a black dashed circle marking the $R_{\rm 200m}$ of the cluster.
    \emph{Right panel:} \text{Shock boundaries overplotted on the Mach number of the gas, as detected by a shock finder}. The complex morphology of the various shocks would make it difficult to find a well-defined accretion shock based on Mach numbers alone.
    }
    \label{fig:shock_algorithm}
\end{figure}

The goal of this work is to find the outer accretion shock when cold gas falls into the hot halo. At the location of the shock-front, infalling kinetic energy is converted to internal energy and entropy, creating a localised rise in these quantities. However, since clusters are inherently aspherical, with gas primarily entering through the filamentary direction, the location of the shock can change as a function of direction. Figure~\ref{fig:shock_algorithm} shows a projection of the gas entropy density (left panel) and the dark matter density (middle panel). The dotted lines in both panels correspond to the ``shock'' surface projected onto this plane. We describe below the algorithm to determine the location of the shock in each direction.

\begin{figure}[t] 
    \centering
    \includegraphics[width=1\columnwidth]{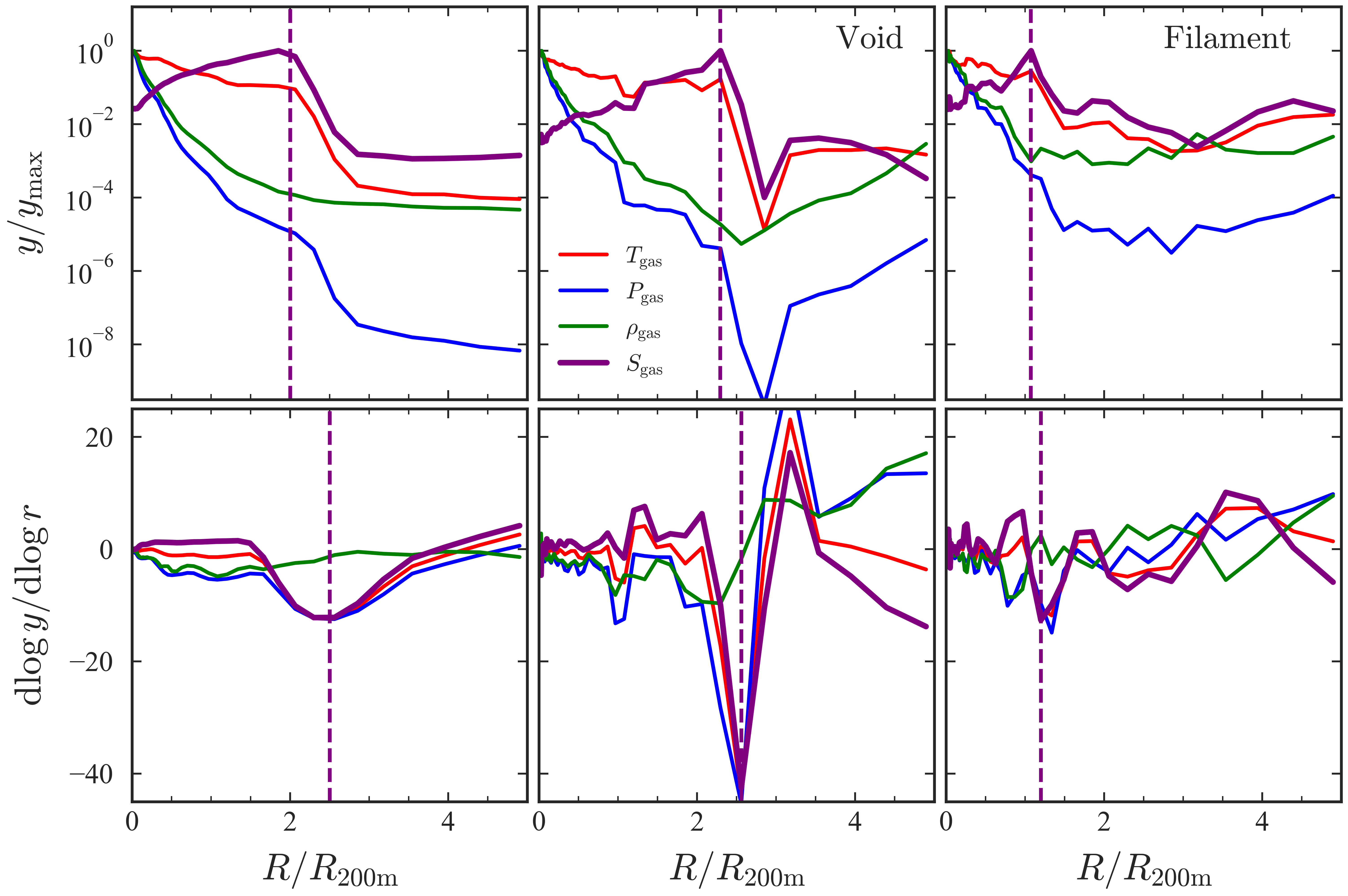} 
    \caption{    
    Radial profiles (\emph{top row}) and their logarithmic slopes (\emph{bottom row}) for gas properties in the same individual cluster from Figure~\ref{fig:shock_algorithm}. The columns compare the angular median profiles (left) with profiles along a representative void (middle) and filamentary (right) direction. The specific sightlines for the void and filament are identified by blue and black arrows, respectively, in Figure~\ref{fig:shock_algorithm}. The vertical dashed lines correspond to the location of shock from entropy maximum (\emph{top row}) and minimum of entropy slope (\emph{bottom row}).
    }
    \label{fig:352_profiles}
\end{figure}

\textit{Shock tracers}: Firstly, we find the best tracer for the shock in the gas distributed around the halo. We compute the pressure, entropy, temperature, and gas density in \(50 \times 50\) solid angle bins, uniformly spaced between \(0<\cos\theta<1\) and \(0<\phi<2\pi\), and use 50 logarithmically-spaced radial bins from $0.1$~Mpc to $20$~Mpc to measure the radial profile of these quantities. The gas temperature is derived from the abundance of electrons ($x_{e}$) and the internal energy ($u$) and is given by,
\begin{equation}
    T = 10^{10} \frac{(\gamma - 1 )u\mu}{k_B}
\end{equation}
where $\gamma = 5/3$ is the adiabatic index of the gas, $k_B$ is the Boltzmann constant in CGS units, $\mu = 4/(1+3X_{H}+4X_{H}x_{e})$ is the mean molecular weight, and $X_{H} = 0.76$ is the hydrogen mass fraction. The gas entropy and pressure are defined as,
\begin{eqnarray}
    S & = &  \frac{T}{\rho^{\gamma -1}} \\
    P & = & n_e k_BT
\end{eqnarray}
where $n_{e} = x_{e}n_{H}$ is the electron number density, and $n_{H}=(X_{H}/m_p)\rho$ is the hydrogen number density.
In Figure~\ref{fig:352_profiles}, we show how the different gas parameters, $\rho, T, S$ and $P$, vary as a function of radius for a single halo (the same halo shown in Figure~\ref{fig:shock_algorithm}). We show the angular median profiles (left panel) and the profiles along two specific directions corresponding to a void (middle panel) and a filament (right panel). Note that the profile is computed in each angular bin; the median corresponds to the median of the angular distribution of each quantity (i.e., $\rho,T,S$ or $P$) at a given $r$. The bottom panel shows the logarithmic-slopes of the profile.

We identify the accretion shock by tracing features in thermodynamic gas profiles near the halo boundary ($\sim 2R_{\rm 200m}$). While the shock signature is visible in temperature, pressure, and density, we find that entropy ($S$) provides the most robust tracer. The physical reason for this is clear: as gas crosses the shock front, its temperature rapidly increases and then remains nearly constant at the virial value. However, its density continues to rise towards the cluster centre. This combination causes the entropy ($S \propto T/\rho^{2/3}$) , which is related to the ratio of temperature and density, to exhibit a distinct maximum precisely at the shock location. We therefore define the shock radius, $R_{\rm sh}$, as the location of this entropy maximum. This method is effective even in challenging environments; the feature is sharp along cold, low-density void directions and remains identifiable even along noisy, substructure-rich filaments. 

An alternative approach, employed by \cite{Aung:2020grs}, identifies the shock via its steepest negative gradient in entropy. We compare the shock locations found using these two tracers: the entropy maximum versus the steepest negative gradient. The offset between these locations is indicative of the width of the gas boundary. Both definitions work as valid choices, with the steepest-gradient method tracing a slightly larger radius. The key difference is that derivatives are often noisy along filamentary directions, whereas the entropy maximum provides a cleaner signal in these dense environments. We therefore adopt the entropy maximum to trace the outer gas boundary, which provides a clean separation of the hot halo gas from the outer region and defines the inner edge of the accretion shock.

\textit{Shock surface}: To construct the 3D shock surface, we first generate entropy profiles along $50 \times 50$ radial sightlines (bins in $\theta, \phi$). Along each sightline, the accretion shock manifests as a distinct entropy maximum, as seen in Figure~\ref{fig:shock_algorithm}. To distinguish these physical shocks from numerical noise, we impose a strict identification criterion: a feature is only classified as a shock if the entropy at its peak ($S_{\rm max}$) is at least twice that of the preceding local minimum ($S_{\rm max}/S_{\rm min} > 2$). This threshold is physically motivated; via the Rankine-Hugoniot jump conditions, an entropy jump of this magnitude naturally selects for shocks with a Mach number $\mathcal{M} \gtrsim 3$. While pristine accretion shocks along voids can have very high Mach numbers ($\mathcal{M} \gg 10$), shocks along pre-heated filaments are weaker, with $\mathcal{M} \gtrsim 2$ \cite{Miniati_2000, 2003ApJ...593..599R}. Our chosen threshold is, therefore, a deliberate strategy: it is stringent enough to discard noise but inclusive enough to capture the weaker yet physically important shocks along filaments. This allows us to build a complete map of the gas accretion boundary, in contrast to studies that focus exclusively on high-Mach-number pristine shocks \cite{Molnar_2009,Valles-Perez:2024lun}.

Using the directional data, we also measure the angular median entropy profile. The median is naturally less sensitive to substructure and shows a sharp entropy jump at the gas boundary. The location of the maximum of this jump is assigned to be the shock radius. 

Optimally, searching for turning points in any distribution is naturally prone to noise, particularly when using directional data that contain only a fraction of particles in each wedge. Additionally, the presence of substructures can create local maxima that do not correspond to the transition into the cluster halo. To distinguish the accretion shock of the halo from those around infalling subhalos, which have a lower virial temperature, we impose a virial temperature-based selection after all maxima have been identified in the entropy profile, in addition to the shock jump conditions. To ensure our shock-finding does not misidentify substructure boundaries, we require the smoothed gas temperature at a shock candidate's location, $T(r=R_{\rm sh})$, to be greater than $0.3~T_{\rm halo}$. The reference temperature $T_{\rm halo}$ is defined as the temperature at $R_{\rm sh,med}$, the shock radius derived from the median gas entropy profile.

The left panel of Figure~\ref{fig:shock_algorithm} shows a colormap of the $x-y$ projection of the gas entropy profile. The colour corresponds to the median entropy in each cell. In this figure, the projection length is taken to be $0.7$~Mpc. The black points correspond to the locations of the maximum entropy jump in each direction. The red curve is a 2D smoothed fit to the black scatter points. The orange curve is the shock boundary found from the logarithmic slope of entropy. The blue and black arrows correspond to void and filamentary directions, the entropy of which is shown in the middle and right panels of Figure~\ref{fig:352_profiles}. The middle panel shows the same shock boundary overplotted on DM density. 

We note that the shock surface is not spherical and is typically much larger than the $R_{200 \rm m}$. We note that the shock radius is typically shorter along the filamentary direction, whereas along the voids, it has traveled to larger distances. This is more evident when we plot the shock surface over the dark matter density, rather than the gas entropy, in the middle panel of Figure~\ref{fig:shock_algorithm}. The right panel of Figure~\ref{fig:shock_algorithm} shows a comparison with the Mach number ($\mathcal{M}>2$) of the shocked gas cells from IllustrisTNG. The outer shock surfaces coincide with the higher Mach number cells, typically of $\sim 100$ and $\sim 4$ along voids and filaments respectively, agreeing with the Illustris-TNG shock finders \cite{Schaal:2016szc}. Additionally, we find that along the filamentary directions, the Mach number fields alone are insufficient to identify a contiguous shock surface; using the entropy based estimator allows us to disentangle this cleanly.

\subsection{Splashback finding algorithm}
\label{sec:splash_algorithm}

We also find the splashback radius in the dark matter distribution as a function of direction. For this work, we use a novel method to extract the splashback radius in each direction. Our method, similar to the SHELLFISH algorithm \citep{Mansfield:2017}, finds the particle distribution in each direction, but additionally uses velocity information to extract the splashback radius. In particular, we analyze the phase space distribution and use the dispersion transition in the radial velocity, rather than the density minimum, to find splashback. A comprehensive description of the methodology and its scientific implications will be presented in the accompanying paper \cite{Sen2025}.  As in Section \ref{sec:shock_algorithm}, we define a grid of \( 50 \times 50 \) solid angle bins, uniformly spaced in \(\cos\theta\) and \(\phi\), along with 100 logarithmic radial bins spanning from 0.1~Mpc to 20~Mpc. We compute the velocity dispersion of the radial velocity component ($\sigma_r$) and its logarithmic derivative along every direction in \(\cos\theta\) and \(\phi\), and we denote the location of the minimum  in the velocity dispersion derivative as the location of the outer boundary of dark matter, or the splashback distance $R_{\rm sp}(\theta)$. 

To ensure that local minima do not trace the change in dispersion around subhalos, we calculate the average dispersion over inner $5$ radial bins from the center of the host halo, $\sigma_{\rm r,halo}$, and use $0.1 \sigma_{\rm r, halo}$ as a dispersion threshold along each sightline; in this sense, we only take those minima in the derivative profile that lie in a region with a dispersion higher than the dispersion threshold. Additionally, we impose an inner radial cut at $0.85R_{\rm sp, med}$ and an outer radial cut at $2R_{\rm sp, med}$, where $R_{\rm sp, med}$ is the splashback found from the logarithmic derivative of the median density profile. In cases where no significant minima are found in the derivative of the dispersion profile, we relax the inner radial cut to $0.6R_{\rm sp, med}$.\\

\begin{figure}[t] 
    \centering
    \includegraphics[width=1\columnwidth]{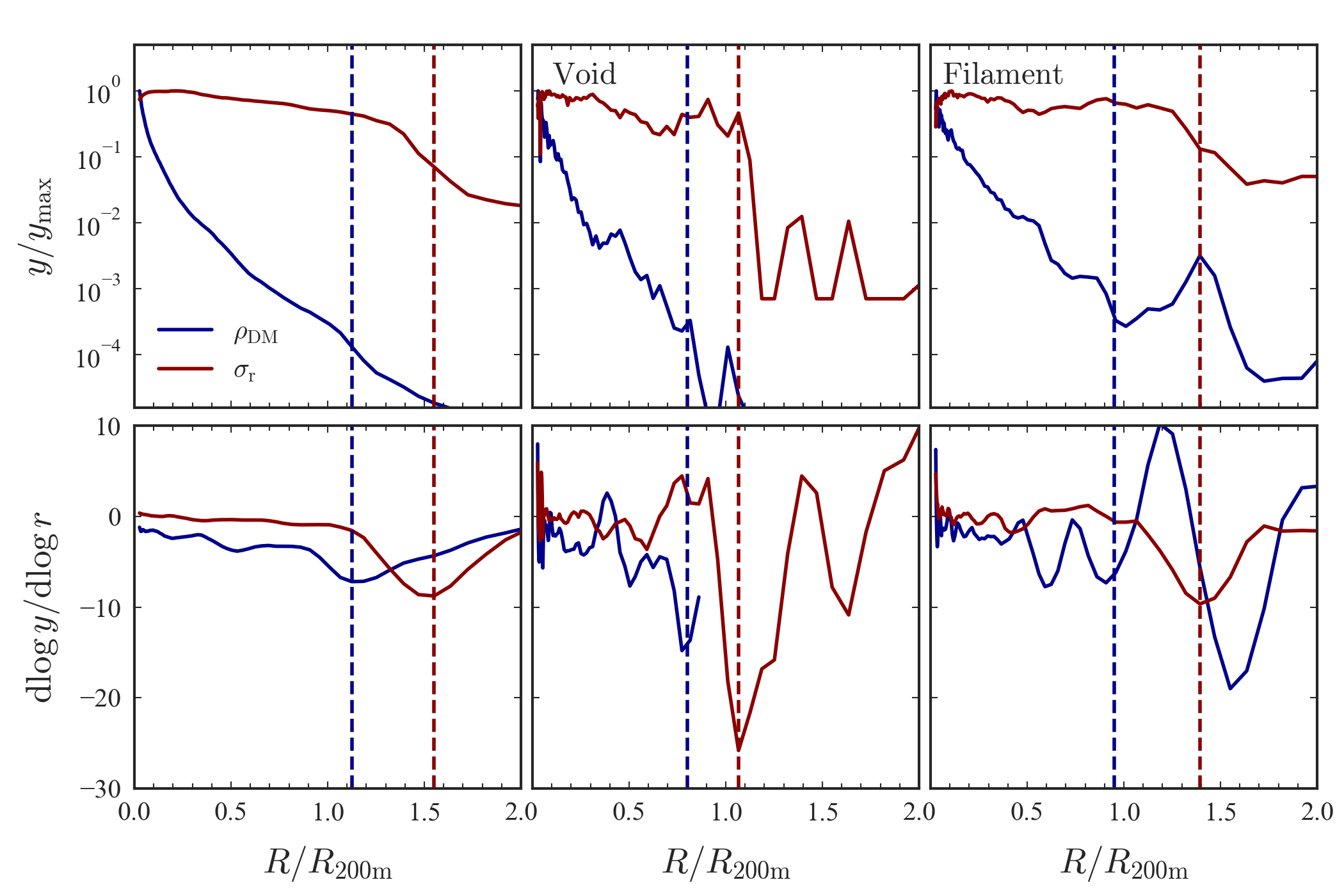}
    \caption{The radial profile (\textit{top panel}) and its logarithmic-derivative (\textit{bottom panel}) of dark matter density (dark blue) and radial velocity dispersion (red) for the same cluster shown in  Figure ~\ref{fig:shock_algorithm} and Figure ~\ref{fig:352_profiles}. The left panel corresponds to the angular median profiles, the middle panel corresponds to profiles along a void direction (blue arrow in Figure~\ref{fig:splashback_algorithm}) and the right panel corresponds to a filamentary direction (black arrow in Figure~\ref{fig:splashback_algorithm}). The vertical dashed lines correspond to the location of minimum of the slope.}
    \label{fig:dm_tracers}
\end{figure}

Figure~\ref{fig:dm_tracers} shows the density and dispersion as functions of radius (top panels) for the cluster shown in Figure~\ref{fig:shock_algorithm} and ~\ref{fig:352_profiles}, as well as their logarithmic slopes (bottom panels). The leftmost panel shows the angular median profile. This corresponds to the median of the values in all angular bins at a given $r$. The middle panel corresponds to the direction of a void, and the right panel corresponds to a filamentary direction. 

In the median profiles, a transition is clearly visible in both the dispersion and density profiles. However, the minimum of the slope occurs at a smaller radius for the density than for the dispersion. This is expected, as the dispersion transition traces the outermost boundary of the velocity phase space, whereas the density minimum traces the point of maximum density, where $v_r \simeq 0$. This point is therefore shifted to a smaller radial distance compared to the full radial extent of the multi-streaming region \cite{Bertschinger:1985pd, FG84, Contigiani:2018qxn}. We find that, on average, the splashback radius determined from the steepest negative gradient of velocity dispersion is $1.2$ times larger than that derived from density. Both quantities are robust probes of the dynamics of halo outskirts. The density-derived minimum is relatively more accessible observationally in the absence of velocity information; however, projected dispersion profiles can be measured in redshift surveys using galaxies as dark matter tracers \cite{Aung:2022buy}.

The robustness of any directional tracer is challenged by low particle statistics and complex environments, such as filaments. Even along clean, high-contrast void directions, where both methods perform reasonably well, the density profile already shows greater susceptibility to fluctuations from substructure. This distinction becomes critical along filamentary directions. Here, the velocity dispersion provides a far superior tracer because the hot, kinematically active halo ($\sigma_r \gg 0$) presents a sharp contrast to the dynamically cold material in an infalling filament or subhalo. This high-contrast kinematic transition is difficult to mimic.

\begin{figure}[t]
    \hspace*{-0.05\linewidth} 
    \includegraphics[width=1.\columnwidth]{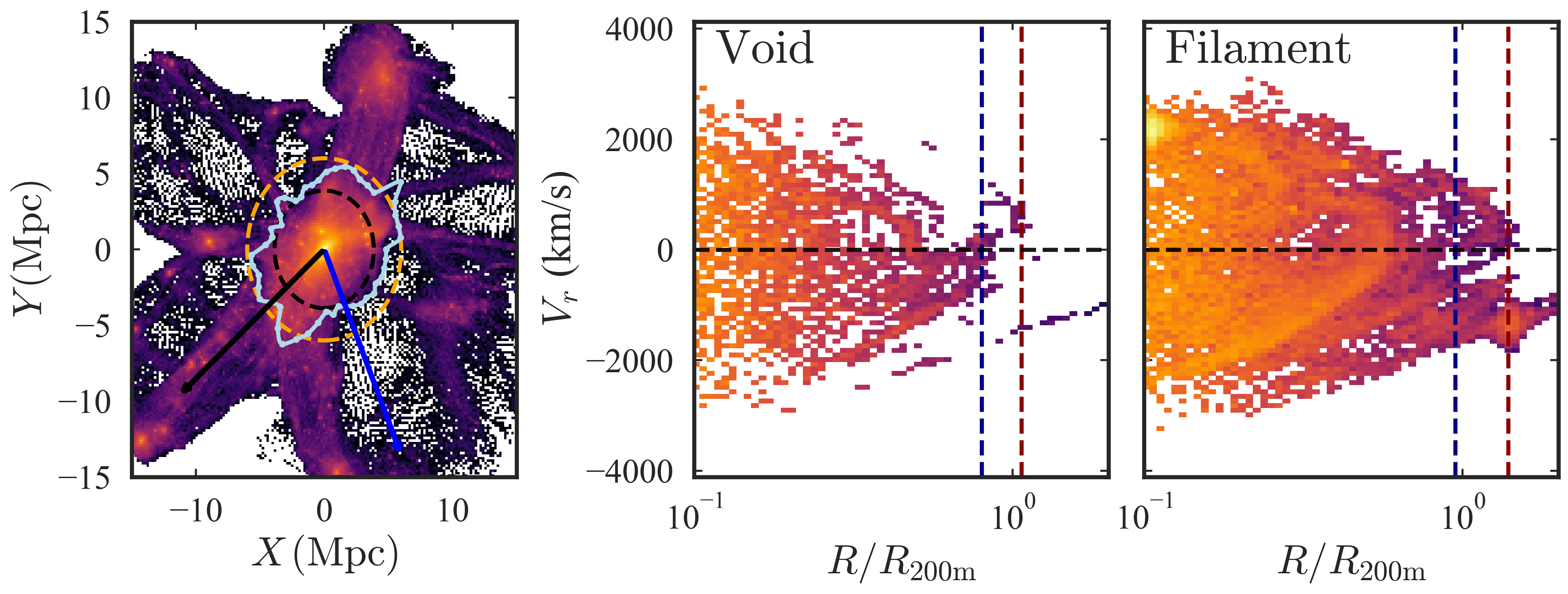} 
    \caption{A 2D projection of the splashback boundary found by our algorithm for the same cluster shown in   Figures~\ref{fig:shock_algorithm}-\ref{fig:dm_tracers}.
    \emph{Left panel:} Splashback boundary over-plotted on a slice of the DM density. The orange and black dashed circles are the spherical estimate of splashback from the logarithmic derivative of the median dispersion profile and $R_{\rm 200m}$ of the cluster, respectively.
    \emph{Middle panel:} The phase space density ($\mathcal{N}$) along the void indicated by the blue arrow in the left panel. 
    \emph{Right panel:} The phase space density ($\mathcal{N}$) along the filament indicated by the black arrow in the left panel. The blue and red vertical line corresponds to the estimated splashback from density and our dispersion-based algorithm.}
    \label{fig:splashback_algorithm}
\end{figure}

Conversely, the density-based method is easily confused in these environments. Since density is effectively a particle count, a compact subhalo just outside the splashback radius can have a local density comparable to the halo's outer regions, creating false minima in the density slope that are unrelated to the true boundary. The dispersion method proves to be a more robust tracer, as shown in the filamentary direction of Figure~\ref{fig:splashback_algorithm} (right panel). There, the dispersion slope (red) presents a clean, unambiguous minimum at the halo boundary, while the corresponding density slope (blue) appears noisy and fragmented by substructure-induced minima. Therefore, as a tracer of the boundary in individual directions, the kinematic dispersion is a fundamentally more reliable probe.

Figure~\ref{fig:splashback_algorithm} demonstrates the implementation of the splashback algorithm on a single halo (the same halo that was used in the previous section). The left panel shows a $0.7$~Mpc thick slice of the projected density in the $x-y$ plane. The light blue points correspond to the splashback in each direction, and the light blue curve is a 2D smoothed fit to the splashback points. The middle panel shows the phase space distribution along an angular bin centered on the blue arrow shown in the left panel, corresponding to a void direction. The red and blue dashed vertical lines in the phase space diagram correspond to the locations of the dispersion and density-slope minima. The right panel shows the phase space distribution along the angular bin centered on the black arrow shown in the left panel. We find that the splashback detection is much cleaner than shock-finding along both the filamentary and void directions. We use this algorithm to detect splashback for all  812 combined clusters from TNG300-1 and TNG-Cluster, covering the entire cluster mass range.

Note that the density-derived splashback radius $R_{\rm sp,dens}$ corresponds to the 75--87th percentile of particle apocenters \cite{Diemer:2017ecy}. By extension, $R_{\rm sp,disp}$ likely represents a comparable but distinct percentile of apocentric distances. A detailed analysis of the particle populations associated with these different splashback definitions will be presented in our companion paper \cite{Sen2025}. Consequently, the measured offset between accretion shock and splashback radii depends critically on the specific operational definitions employed for each boundary. In addition, it will also be interesting to explore the relationship between the dispersion transition and the edge radius \citep{Garcia:2022zsz}, as well as accretion based boundaries such as the depletion radius \cite{Fong:2020fuf}.

In the next section, we discuss our results on the statistics of the distribution of splashback and shock, as well as the interconnections between these surfaces around cluster mass halos.

\section{Results}
\label{sec:results}

\subsection{Angular Distribution of shock and splashback}

\begin{figure}[t]
    \centering
    \includegraphics[width=0.77\columnwidth]
    {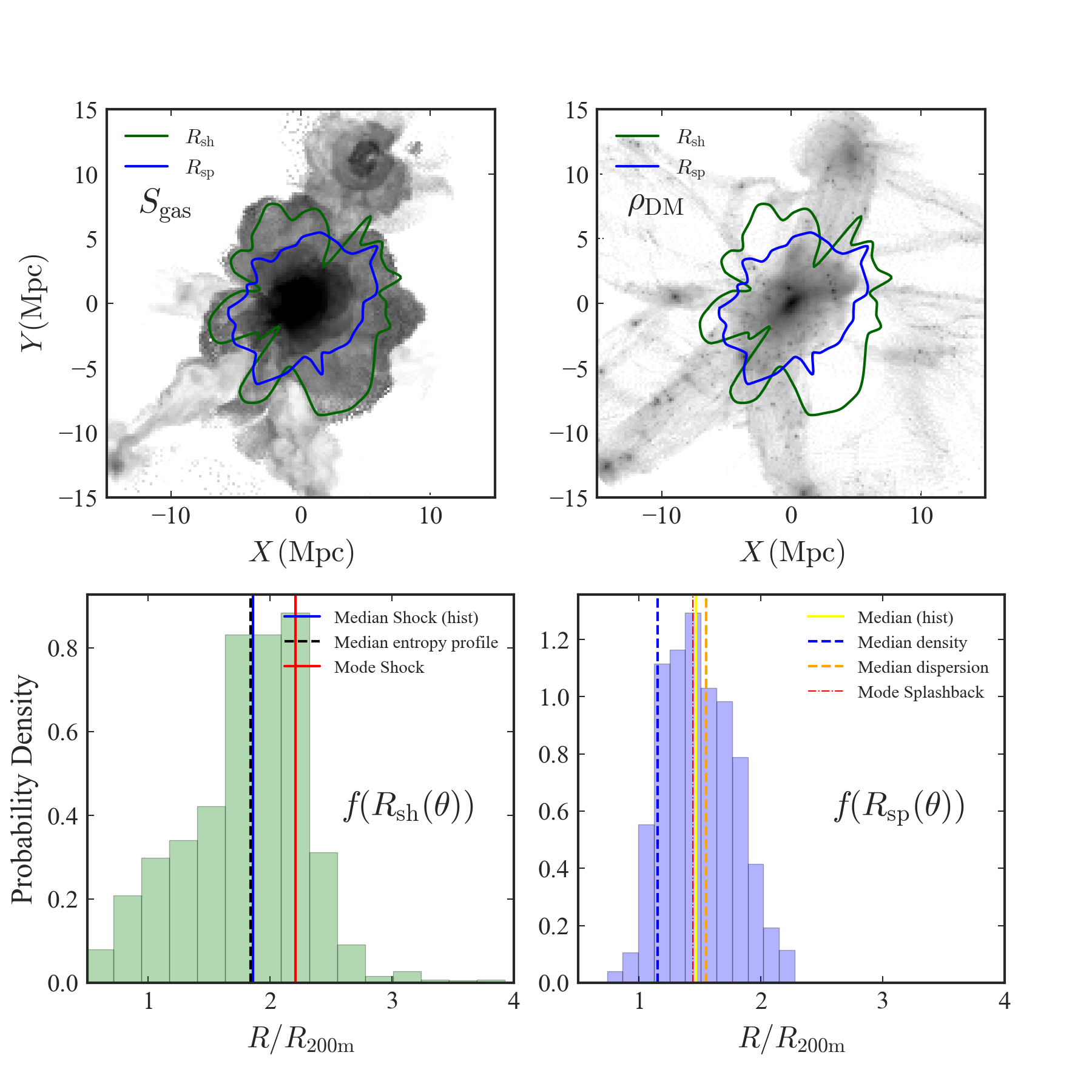}
    \caption{
    An example of the shock and splashback boundaries in a massive galaxy cluster shown in Figure~\ref{fig:shock_algorithm}.
    \emph{Top panel}: Maps of gas entropy (left) and dark matter density (right). The identified splashback radius (blue) and accretion shock radius (green) are overlaid.
    \emph{Bottom panel}: Angular distribution of the splashback radius (right) and shock radius (left). For this halo, the median and mode of each distribution agree to within 5\%.
    }
    \label{fig:shock_splash_sidebyside}
\end{figure}

The mass distribution of dark matter halos is typically not expected to be spherical \cite{Jing:2002np}; therefore, both the shock and splashback surfaces are also expected to be aspherical. Rather than characterizing these surfaces by a single radial value, it is instructive to examine the distributions of the splashback and shock radii as functions of direction. The top panel of Figure~\ref{fig:shock_splash_sidebyside} shows the shock and splashback surfaces overlaid on a 2D slice of the gas entropy colormap (left panel) and the dark matter density (right panel). The distributions $f(R_{\rm sh}(\theta))$ and $f(R_{\rm sp}(\theta))$, obtained using the algorithms described in the previous sections for a single cluster, are shown in the bottom panel. 

We first observe that the splashback and shock surfaces do not coincide; the gas typically extends much farther out. We find that this difference arises particularly along void directions, as seen in the top-right panel where the surfaces are overlaid on the dark matter density (for which the filamentary structure is more clearly visible). Along filamentary directions, the shock radius is typically pushed inward and becomes more consistent with the splashback location. Overall, the splashback surface is significantly more regular, with a relatively uniform curvature, whereas the shock surface changes curvature sign as one moves along it.

Regarding the distribution of directional radii, we find that the median and mode agree well for splashback, but are offset for the shock, indicating tails at smaller radii in the filamentary direction. The shock radius exhibits significantly greater directional scatter compared to the splashback radius overall. This trend is consistently observed across all clusters in the simulation. For the splashback radius, the median of the distribution along void and filamentary directions does not differ dramatically, consistent with the findings of \cite{2022A&A...667A..99W}, who report a nearly isotropic distribution when accurately capturing the outermost caustic. However, the distribution of dark matter and gas along voids and filaments can potentially encode a wealth of information about triaxiality and accretion history \citep{2011ApJ...734..100L, Adhikari:2014lna, 2022A&A...667A..99W} that merits more detailed investigation in future work within this framework.

Given the distributions, to define a single representative radius, we use the median of each distribution, which provides a typical characterization of the gas or dark matter boundary in different directions. In the next section, we compare the full distribution of these medians across the cluster sample to understand the basic trends of these boundaries with halo properties.

\subsection{Statistical variation across halos}

As discussed above, we use the median of the angular distribution of the shock and splashback radii as the nominal “radius’’ defining $R_{\rm sp}$ or $R_{\rm sh}$. We adopt the median primarily because the mean can be biased by local overdensities; moreover, the angular distributions are asymmetric, as seen in the bottom panel of Figure~\ref{fig:shock_splash_sidebyside}. One could alternatively use the mode; however, since the median and mode agree well for most cluster-mass halos, we present our fiducial results using the median.\footnote{We have also used the spherical radius corresponding to the volume enclosed by the splashback and shock surfaces, following \cite{Aung:2020grs}, and find consistent results.}

\begin{figure}[t] 
    \includegraphics[width=1\columnwidth]{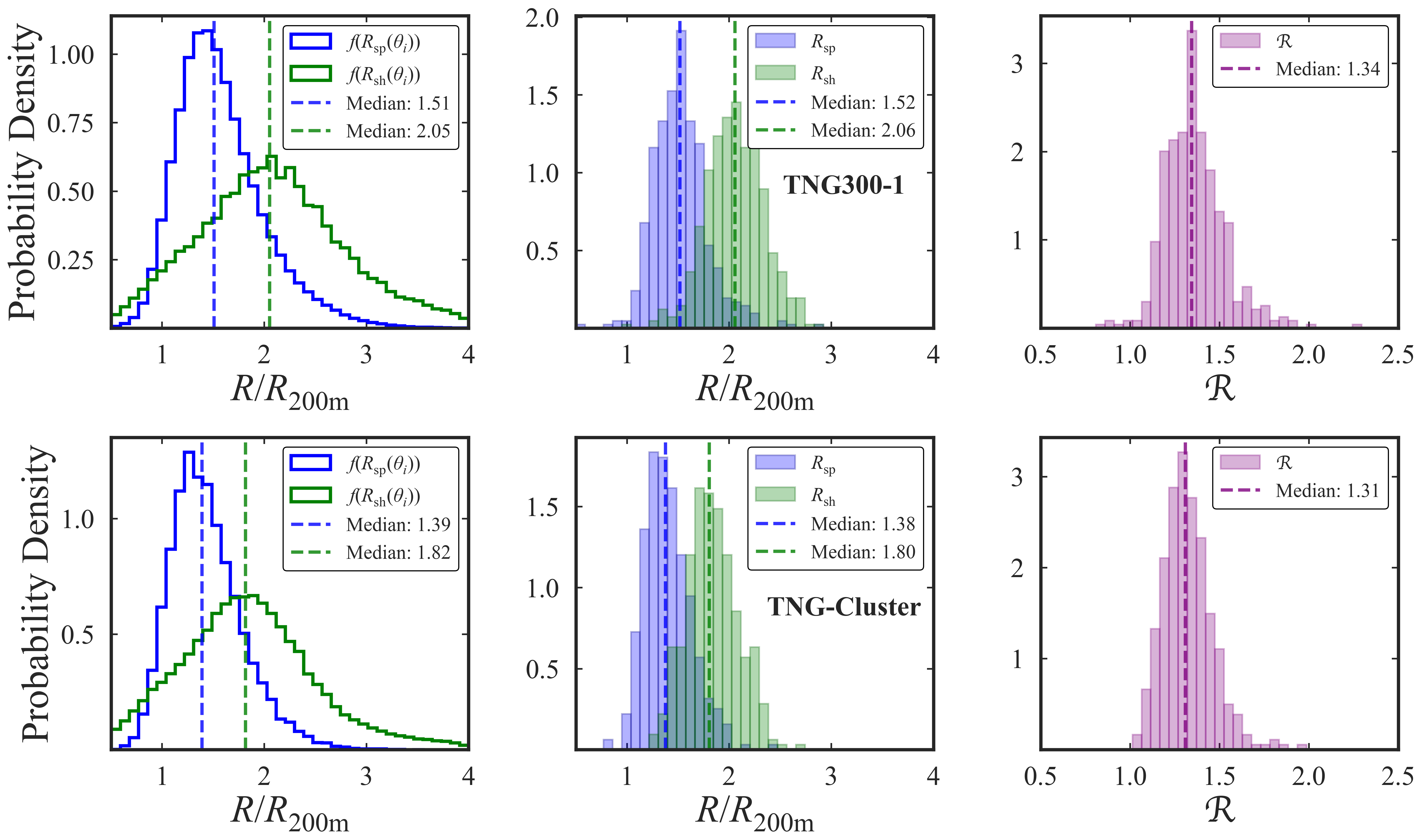} 
    \caption{\emph{Left panel:} The total angular distribution of splashback (blue) and shock (green) for all the clusters in TNG300 (\emph{top}) and TNG-Cluster (\emph{bottom}) samples. \emph{Middle panel:} Distribution of median values from the angular distributions for splashback (blue) and shock (green).\emph{Right panel:} Distribution of ratios ($\mathcal{R}\equiv R_{\rm sh}/R_{\rm sp}$) is calculated using the medians from middle panel. Dashed vertical line shows the median values in every panel.}
    \label{fig:stat analysis}
\end{figure}

The top row of Figure~\ref{fig:stat analysis} shows the distributions of the splashback and shock radii for clusters from TNG300-1, while the bottom row presents clusters from TNG-Cluster. These include 352 clusters from the TNG-Cluster simulation and 460 clusters from the TNG-300 box with $M_{\rm 200m}>10^{14} M_\odot$. The left panel displays the net angular distributions of splashback (blue) and shock (green) radii for the combined directional measurements from all clusters. The middle panel shows the distribution of the median value for each individual cluster. Consistent with previous findings, we observe a clear offset between the splashback and shock radii, with the shock radius typically extending farther out.

The rightmost panel depicts the distribution of ratios ($\mathcal{R}\equiv R_{\rm sh}/R_{\rm sp}$) computed from the median values for each cluster. We find an average ratio of $1.34^{+0.17}_{-0.13}$ for TNG300-1 and $1.31^{+0.14}_{-0.12}$ for TNG-Cluster, as indicated by the dashed vertical line. This ratio is smaller than the $1.89\pm0.16$ reported by \cite{Aung:2020grs}, primarily due to our use of the dispersion-based splashback estimator, which generally yields larger radii than the density-slope–minimum estimator. Additionally, our choice of using entropy maxima rather than the minimum of the logarithmic derivative of entropy contributes to this difference. The dispersion-based definition more accurately captures the phase-space “boundary,” encompassing the multistreaming, high-dispersion region of the halo by construction.

For comparison, we also performed an analysis using the traditional method for measuring splashback by taking the minimum of the logarithmic derivative of dark matter density. Given the noisiness of the density derivative in individual directions, we instead used the minima of the median dark matter density profile (similar to the first column of Figure~\ref{fig:dm_tracers}) for the splashback radius and the minimum of the logarithmic derivative of the angular median entropy profile as a tracer for the shock radii (similar to \cite{Aung:2020grs}). We computed these quantities for each halo in the combined sample and found an offset of $2.10^{+0.39}_{-0.26}$, consistent with the previous work \cite{Aung:2020grs,Zhang:2024tue}. Because our fiducial method identifies the gas boundary using entropy maxima interior to the slope minima and dark matter boundaries using dispersion minima exterior to the density minima, our algorithm provides a lower limit on the offset between the two boundaries. Nonetheless, even with this more inclusive estimate of the dark matter boundary, we still observe a persistent offset between the splashback and shock radii, indicating that the displacement between the two features is a robust effect.

Overall, the scatter in the shock radius is significantly larger than that of the splashback radius, a trend evident in both the full angular distributions and the cluster-by-cluster median distributions. This behaviour reflects the strong directional dependence of the shock location, which varies significantly between voids and filaments. 

Along the filamentary directions, the shock radius is typically pushed inward and is closer to the estimated splashback radius. We also find the median of the shock distribution only along void directions. To remove the filamentary directions, we note that along the void directions, external accretion shocks typically have high mach numbers ($\mathcal{M}>10$), so we mask all the directions where the ratio of entropy becomes less than 50 (i.e., $S_{\rm max}/ S_{\rm min} < 50 $) or, equivalently, where the Mach number is less than 20 ($\mathcal{M}<20$) in that cluster.  In this case, the median ratio is $1.47^{+0.19}_{-0.15}$ for TNG300-1 and $1.42^{+0.17}_{-0.12}$ for TNG-Cluster. We find an average filament coverage of around $35\%$. For robustness, we repeated our analysis according to \cite{Zhang:2025eyu}, i.e., using a temperature-based cut of about $10^{5}$~K to classify into filament or void directions, and found similar results. Along the filamentary directions, the median ratio is $1.00^{+0.13}_{-0.12}$ and $1.03^{+0.16}_{-0.13}$ for TNG300-1 and TNG-Cluster, respectively.

A study of resolution effects on the boundaries is presented in Appendix A. Firstly, we separate the combined sample of TNG300-1 and the TNG-Cluster zoom-in suite since the halo mass functions differ between these two simulation suites and verify that our results are consistent between the two simulations. We find that the median ratio measured from the TNG300-1 clusters is in agreement with our findings from the TNG-Clusters. Additionally, to test the resolution dependence of our algorithm, we select 100 clusters above $10^{14} M_\odot$ from different resolution runs of TNG300 (TNG300-1, TNG300-2, and TNG300-3) and compare the distributions of shock and splashback radii between them. We find that the splashback estimate is very robust against resolution; however, the shock requires high resolution for robust detection.

In the next section, we study the impact of various halo properties, such as halo mass, accretion rate, etc., on the offset in detail and the individual boundary radii.

\subsection{Halo properties and the ratio of splashback and shock}

\begin{figure}[t]
    \centering
    \includegraphics[width=1\textwidth]{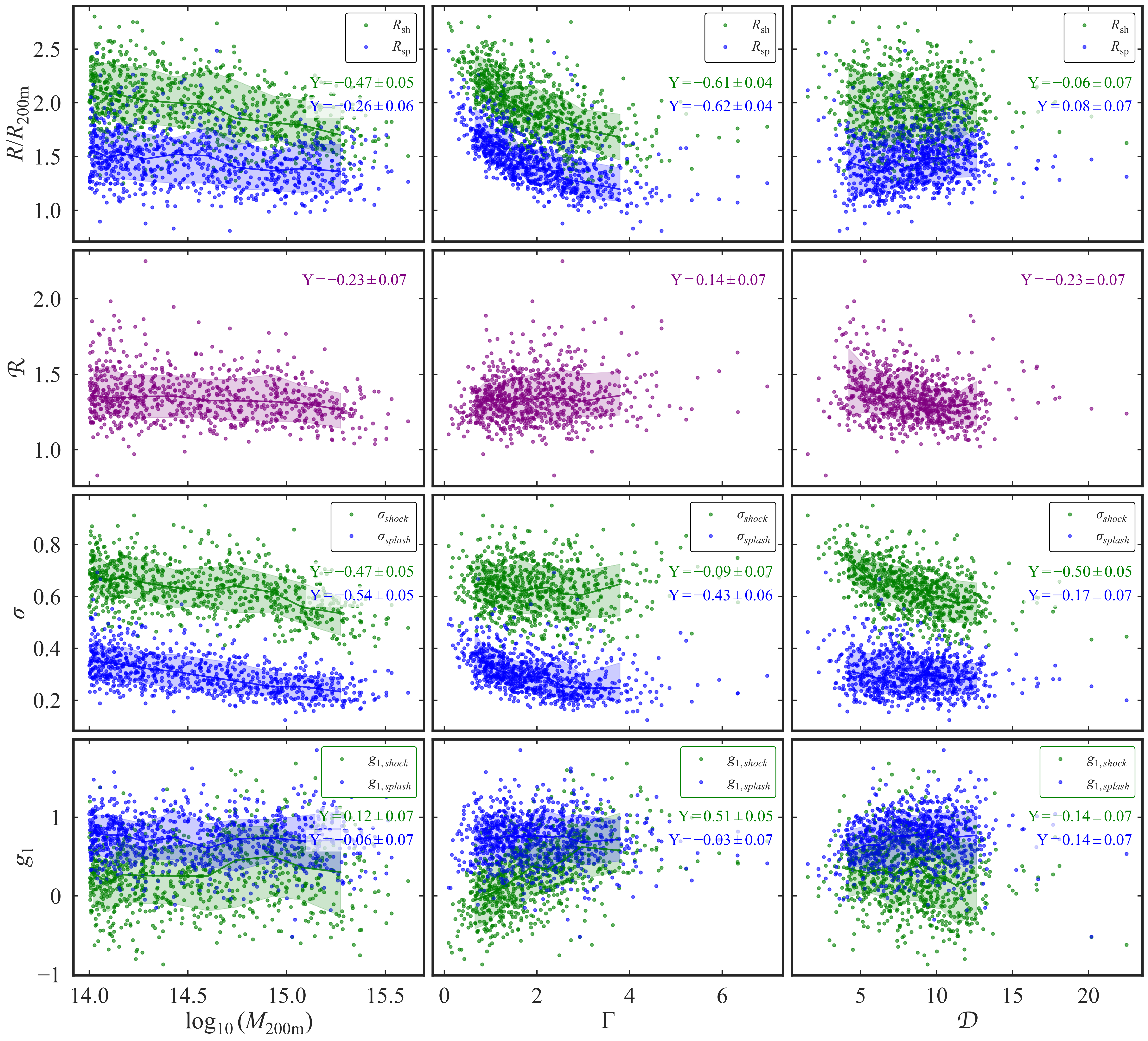}
    \caption{
    Correlations of shock (green) and splashback (blue) boundary properties with halo properties. Each column investigates a different halo property: halo mass ($M_{\rm 200m}$), accretion rate ($\Gamma$), and dispersion slope depth ($\mathcal{D}$). From top to bottom, the rows examine the normalized median radius ($R/R_{\rm 200m}$), the ratio of radii ($\mathcal{R}\equiv R_{\rm sh}/R_{\rm sp}$), the standard deviation ($\sigma$), and the skewness ($g_1$) of the boundaries angular distributions. Points show individual halos, with the solid line and shaded band representing the median trend and the 16th–84th percentile range. The Spearman correlation coefficient ($\Upsilon$) is reported in each panel.
    }
        \label{fig:scatter_352}
\end{figure}

In this section, we investigate the effects of several relevant properties of the halo with this offset. For this analysis, we use the complete combined sample of cluster mass halos with $M_{\rm 200m}>10^{14}M_{\odot}$ from the TNG300-1 and TNG-Cluster suite. This enhanced sample has 460 halos from TNG300-1 and 352 halos from TNG-Cluster, giving a total of 812 halos covering the entire cluster mass range. This combined dataset provides superior sampling across the full cluster mass range, particularly at both the lower and higher mass extremes. In Figure~\ref{fig:scatter_352}, we show the correlations with halo mass, accretion rate, and the dispersion depth.\\
\\
\textit{Correlation with Halo mass:} Generally, the splashback radius and the shock radius grow with halo mass, similar to the virial radius. Here, we examine the mass dependence of these boundaries in units of $R_{\rm 200m}$. The first column in Figure~\ref{fig:scatter_352} shows the variation of $R_{\rm sh}/R_{\rm 200m}$ and $R_{\rm sp}/R_{\rm 200m}$ and their ratio $\mathcal{R}\equiv R_{\rm sh}/R_{\rm sp}$ with $\log(M_{\rm 200m})$. We observe that the ratio $\mathcal{R}$ exhibits a negative correlation with halo mass, with a Pearson coefficient of $\Upsilon(\mathcal{R}|\log(M_{\rm 200m})) = -0.23\pm0.07$. This indicates that higher-mass halos tend to have lower shock-to-splashback ratios.  The individual variations of $R_{\rm sh}$ and $R_{\rm sp}$, scaled by $R_{200 \rm m}$, can be used to elucidate the underlying causes for this mass dependence.  The individual correlations imply that the dependence of the ratio on mass is primarily driven by the shock radius. While $R_{\rm sp}/R_{\rm 200m}$ shows a negative correlation with halo mass of order $\Upsilon(R_{\rm sp}/R_{\rm 200m})\sim -0.26\pm0.06$, the shock radius in units of $R_{\rm 200m,}$ shows a strong negative correlation with $\Upsilon(R_{\rm sh}/R_{\rm 200m})\sim -0.47\pm0.05$. In Appendix \ref{appendix:resolution_study}, we also show these relations for the TNG-Cluster and TNG300-1 samples separately. We note that the effect is clearer in the more massive, mass-selected TNG-Cluster sample.

This result is somewhat surprising; When scaled by the $R_{\rm 200m}$, theoretically, we do not expect a net correlation of splashback or shock with mass. We expect the location of these boundaries to be largely scale-free, set simply by the accretion rate of the halo \cite{Bertschinger:1985pd, Diemer:2014xya, Adhikari:2014lna}. In particular, we expect that for  a given halo accretion rate, $R_{\rm sp}$ appears at a specific overdensity $\delta_{\rm sp}$, and $R_{\rm sh}$ follows a similar relation. 
For splashback, this has largely been found to be true in simulations, particularly for the traditionally defined splashback radius located at the minimum of the logarithmic slope profile. A mild correlation of the splashback radius with mass has been found using a particle trajectory-based method in SPARTA \cite{Diemer2017}, even when accounting for changing accretion rates.\footnote{SPARTA measures splashback radii for each particle and then for each halo, rather than finding the slope-minimum from halo mass profiles stacked over mass bins.} It is possible that such a correlation may arise in a hierarchical structure formation scenario where multiple mergers take place as the halo grows over time, and accretion is not smooth and spherical as assumed in theoretical models. However, it appears from this work that for splashback, the deviation is not as severe as it is for the gas shock boundary. The deviation of the shock radius from scale-free behaviour is larger compared to the splashback radius and is likely related to the difference in the response of gas and dark matter to mergers during hierarchical structure formation \cite{Zhang:2019kej}.
The shock radius is set by a balance of the inner gravitational pull and the pressure gradient in the shock-heated infalling gas. Beyond the theoretical self-similar models, energetic mechanisms such as radiative cooling (which is not present in dark matter) or pre-heating can lead to deviations from self-similarity. However, the cooling timescales in cluster outskirts are typically much longer than the local dynamical timescale. Therefore, shock heating by mergers or pre-heating before infall is the more likely cause of this deviation.

\begin{figure}[t]
    \centering
    \hspace*{-0.05\linewidth}  
    \includegraphics[width=1\columnwidth]{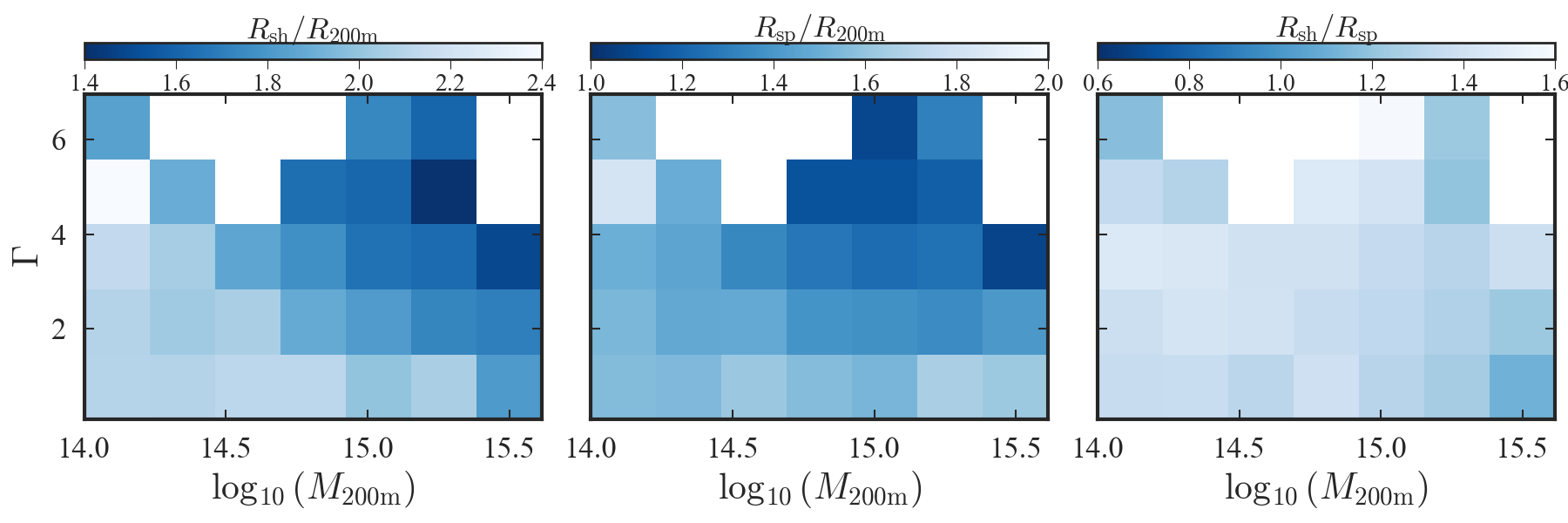}
    \caption{
    The mean shock radius ($R_{\rm sh}/R_{\rm 200m}$, left), splashback radius ($R_{\rm sp}/R_{\rm 200m}$, middle), and their ratio ($\mathcal{R}\equiv R_{\rm sh}/R_{\rm sp}$, right) binned in the 2D plane of halo mass and accretion rate ($\Gamma$). At a fixed mass, both radii decrease with higher accretion rates. 
    However, unlike $R_{\rm sp}$, $R_{\rm sh}$, when normalized with $R_{\rm 200m}$, shows mass dependence when $\Gamma$ is kept fixed, as shown in the left plot.
    }
    \label{fig:2d_correlations}
\end{figure}

To test the robustness of this result and to disentangle any remnant effects of a trend between mass and accretion rate, we check the mass rank correlations with $R/R_{\rm 200m}$. A rank is obtained by a linear map of the halo mass between $0$ and $1$ in bins of accretion rate; the relation between the rank and the boundary radius is then investigated. We find that using mass ranks, the correlation between $R_{\rm sp}/R_{\rm 200m}$ is nearly removed, whereas $R_{\rm sh}/R_{\rm 200m}$ shows a correlation of $-0.27$. The ranking method, however, confounds information in different accretion rate bins; to mitigate this. Figure~\ref{fig:2d_correlations} shows the variation of splashback and shock simultaneously with mass and accretion rate in a 2D plane. We use 7 bins of halo mass and 5 bins of accretion rate and plot the mean $R_{\rm sh}/R_{\rm 200m}$ (left panel) and $R_{\rm sp}/R_{\rm 200m}$ (middle panel). Along the accretion rate axis at a fixed mass, the expected behaviour of splashback and shock radius moving to a higher overdensity ($R/R_{200m}$)  is clearly observed.  Additionally, along a fixed $\Gamma$ direction, we find that $R_{\rm sh}/R_{\rm 200m}$ decreases for higher mass halos. The size of the latter effect is also a function of the accretion rate.

\textit{Correlation with Accretion rate:} To study the dependence on accretion, we adopt the definition of the accretion rate for halos from \cite{Diemer:2017ecy}. We compute
\begin{equation}
    \Gamma_{\rm dyn} = \frac{\log \left[M_{\rm 200m}(t) /  M_{\rm 200m}(t-t_{\rm dyn})\right]}{\log \left[a(t) /  a(t-t_{\rm dyn})\right]}
\end{equation}

where $a$ is the scale factor and $t_{\rm dyn}(z)=2R_{\Delta}/v_{\Delta}$ is the dynamical time of the halo, $v_{\Delta} = \sqrt{GM_{\Delta}/R_{\Delta}}$ at $\Delta=200$, the matter overdensity at which $R_\Delta$ is defined.

The middle column of Figure~\ref{fig:scatter_352} shows the dependence of the splashback and shock radii on the halo accretion rate ($\Gamma$). The accretion rate, along with mass, is a primary determinant of halo boundary size, a relationship that has been extensively studied in dark matter simulations \cite{Diemer:2017uwt,Shin:2022iza}. Consistent with this previous work, we find strong anti-correlations for both boundaries in our hydrodynamic simulations. The Pearson coefficients are $\Upsilon(R_{\rm sp}|\Gamma)=-0.62\pm0.04$ for the splashback radius and $\Upsilon(R_{\rm sh}|\Gamma)=-0.61\pm0.04$ for the shock radius. Separate results for TNG300-1 and TNG-Cluster are shown in Appendix \ref{appendix:resolution_study}.

Our estimates for splashback and shock radius are derived from the median of the angular distributions, as described in Section~\ref{sec:shock_algorithm}. Along with the median, the bottom panel of Figure~\ref{fig:scatter_352} shows the scatter ($\sigma$) and skewness ($g_{1}$) in the angular distributions of the two boundaries in gas and splashback in the distribution of these boundaries. The scatter in the gas shock angular distribution does not correlate significantly with the accretion rate of the halo. The scatter in the angular distribution for splashback, however, has a strong negative correlation ($\Upsilon(\sigma_{\rm sp}|\Gamma)=-0.43\pm0.06$) and a much tighter relation. The spread in the angular distribution is related to the sphericity of the boundary surface; the fact that the scatter in splashback reduces with the accretion rate implies that faster accreting halos tend to be more spherical in nature. The same correlation, however, does not hold for the gas shock boundary. For the shock surface, we find that the skewness of the angular distribution does correlate with the accretion rate with $\Upsilon(g_{1,\rm sh}|\Gamma)=0.51\pm0.05$. This is an effect of the larger accretion rate halos having a larger fraction of their volume in filamentary directions. 

\textit{Correlation with Dispersion depth:} The last column in Figure~\ref{fig:scatter_352} shows the variation of the individual measurements of $R_{\rm sh}/R_{\rm 200m}$ and $R_{\rm sp}/R_{\rm 200m}$, as well as the ratio $\mathcal{R}\equiv R_{\rm sh}/R_{\rm sp}$ with dispersion-slope depth $\mathcal{D}$, defined as the minimum of the logarithmic derivative of the median radial velocity dispersion ($\sigma_{r}$) with respect to the cluster centric radius, $r$. The depth is a tracer of the sharpness of the transition from the inner virialized region of the halo to the outer infall. In particular, it should trace the environmental distribution of the halo. The ratio $\mathcal{R}$ shows an anti-correlation with the depth, with a Pearson coefficient of $\Upsilon(\mathcal{R}|\mathcal{D})=-0.23\pm0.07$. $R_{\rm sh}/R_{\rm 200m}$ and $R_{\rm sp}/R_{\rm 200m}$ show negligible correlations of $-0.06\pm0.07$ and $0.08\pm0.07$, respectively, in the combined sample. However, we note that the mass-selected TNG-Cluster sample behaves differently from the TNG300-1 sample alone. We find a strong negative correlation of $R_{\rm sp}/R_{\rm 200m}$ with $\mathcal{D}$ and $\Upsilon(R_{\rm sp}/R_{\rm 200m})=0.41\pm0.09$ in the TNG-Cluster sample, which we do not see in gas. This difference is likely because the TNG-Cluster sample is more massive than the TNG300-1, which is dominated by halos around $10^{14} M_\odot$. Typically, massive halos dominate their environments, and intrinsic trends are clearer. The trend of the location of the splashback radius with the depth of the feature arises because of the remnant dependence of the accretion rate—the larger the splashback radius, the lower the ambient accretion and, therefore, the velocity dispersion. The shock radius location does not trace the depth of the splashback feature.

\begin{figure}[h] 
    \includegraphics[width=0.9\columnwidth]{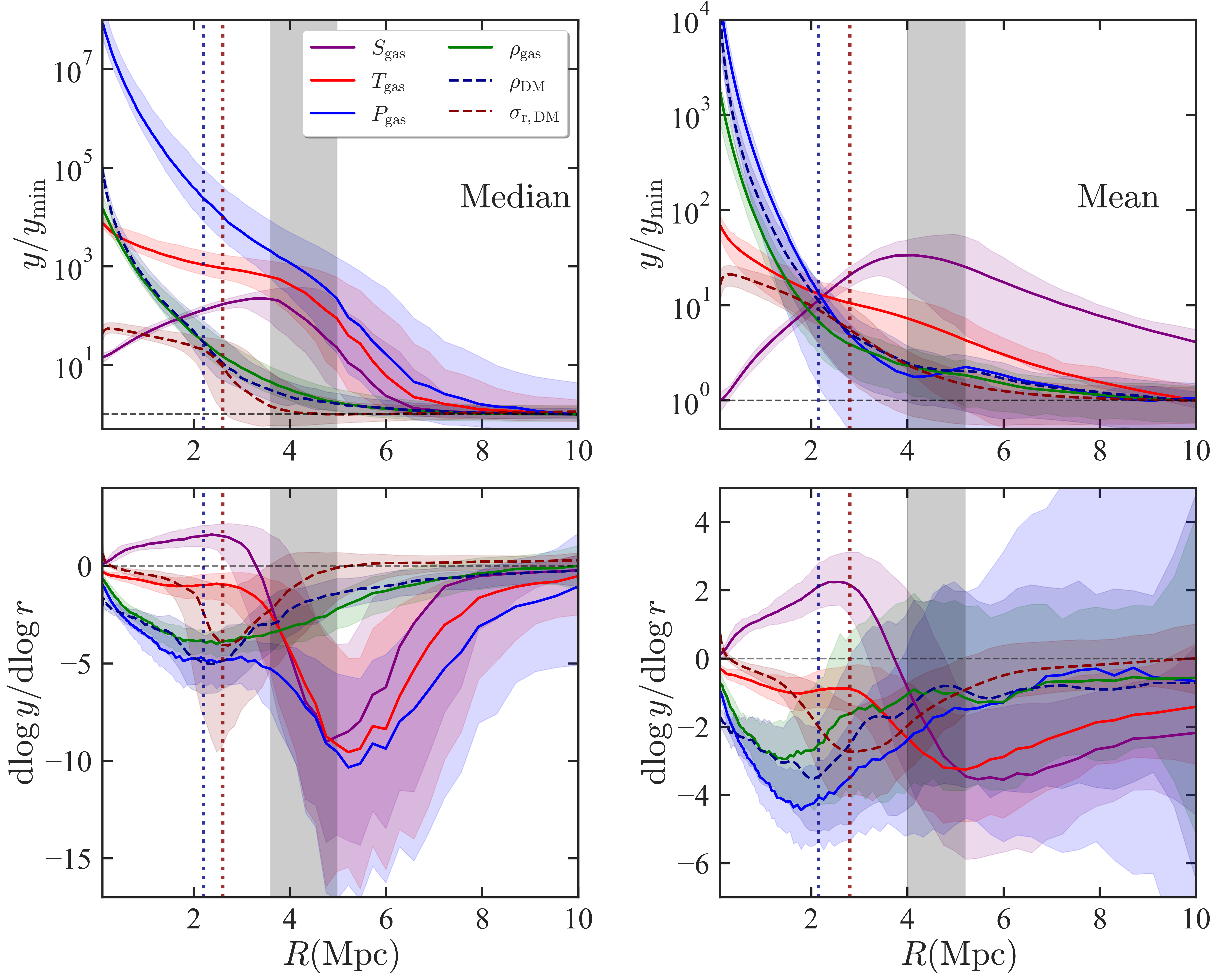}
    \caption{Stacked results for the sample of 460 clusters in TNG300-1. \textit{Top panel}: Angular median (\emph{left}) and mean (\emph{right}) profiles as a function of radius for the different gas and dark matter parameters. \textit{Bottom panel}: Their logarithmic slope profiles as a function of radius for the same set of parameters. The shaded region shows the scatter over the entire cluster sample. The black shaded region indicates the shock region between maximum of entropy and the minimum of its logarithmic slope. Locations of splashback from logarithmic derivative of density (blue) and dispersion (red) are marked by vertical lines.} The sharp features in the pressure and temperature profiles at the outer shock radius are washed out in the angular mean profiles, complicating observational detections (e.g., via the SZ signal).
    \label{fig:352_profiles_1}
\end{figure}

\subsection{Stacked Profiles of Gas and Dark Matter}

So far, we have focused on splashback and shock radii for individual galaxy clusters. Observationally, however, stacked cluster samples are often used to achieve higher signal-to-noise estimates of dark matter and gas distributions. Here, we examine the 3D stacked quantities for our cluster sample. Figure~\ref{fig:352_profiles_1} shows the stacked profiles for all relevant thermodynamic quantities in gas, as well as dark matter density and velocity dispersion for the 460 clusters in TNG300-1. We choose this sample as it represents the true, volume-limited cosmological halo mass function, rather than the mass-selected sample of TNG-Cluster. A volume-limited sample is more representative of observational selection, though additional biases from mass-observable relations are not accounted for in this work. Instead of radial units scaled by $R_{\rm 200m}$, we show results as a function of distance in Mpc.

The top panel shows the stacked angular median (left) and mean (right) profiles. For each halo, we compute the angular median or mean over the angular distribution at each radial bin, then stack these profiles across the full sample. The bottom panel shows their logarithmic derivatives. The shaded regions indicate the total scatter across all halos. For the median profiles, the stacked pressure, entropy, and temperature profiles show a clear signature of the shock in the outskirts, whereas the gas density does not show a prominent dip near the outermost shock—it reaches a minimum closer to the splashback radius. The splashback feature in dark matter dispersion and density is clearly visible in the stacks. In the top right panel, we show the stacked mean profiles. Notably, for the TNG300-1 sample, taking the angular average washes away the sharp outer shock feature in the pressure profile. This occurs because the average is density-weighted, and most of the outer density lies along filaments. In the Appendix \ref{appendix:environment}, we show the stacks separated by void and filament directions. We note that along void directions the logarithmic profile picks up a second outer feature, smearing out the outer profile. For a cluster sample with masses $M>8\times 10^{14} M_\odot h^{-1}$, previous work finds that relaxed clusters show a clear feature in the stacks at the outer shock boundary \cite{Baxter:2021tjr}, consistent with our results (also shown in Appendix \ref{appendix:environment}).

The difference between stacked median and mean pressure profiles has important implications for observing the outer shock feature. The SZ effect is primarily sensitive to the line-of-sight integrated pressure profile of clusters. Typically, angular average stacks are easiest to measure because individual directional profiles are noisy. Our results indicate that these angular average profiles do not accurately capture the outermost gas shock boundary, which may explain the observed alignment of boundaries in SZ and weak lensing measurements \cite{Anbajagane:2021bnx}. Intriguingly, recent observational work finds differences in the measured logarithmic slope profiles along filamentary versus non-filamentary directions \cite{DES:2023laf}, consistent with trends in this work and previous simulations \cite{Aung:2020grs}; however, higher signal-to-noise measurements are required to definitively confirm the external shock. Additionally, our findings indicate that inferences about the behaviour of $n_e$ and $T_{\rm e}$ from spherically averaged pressure profiles should be made with caution. In principle, more information can be gained by measuring profiles along void and filamentary directions separately using multi-probe observations, such as thermal, kinematic, and relativistic SZ effects, as well as forthcoming FRB observations.

\subsection{Redshift Evolution}

To investigate the physical origin, onset, and evolution of the offset and scatter between the shock and splashback radii, we track the evolution of $R_{\rm sh}$, $R_{\rm sp}$, and $R_{\rm 200m}$ in three sample halos selected to represent low, medium, and high mass accretion histories. We use all the publicly available snapshots for the TNG-Cluster simulation. There are 75 redshift bins spaced between $0 < z< 3.01$. We also track the instantaneous accretion rate ($\Gamma_{\rm inst}$) and the accretion rate averaged over the dynamical timescale ($\Gamma_{\rm dyn}$) for each of these clusters in the sample. 

\begin{figure}[t]
    \centering
    \includegraphics[width=0.9\columnwidth]
    {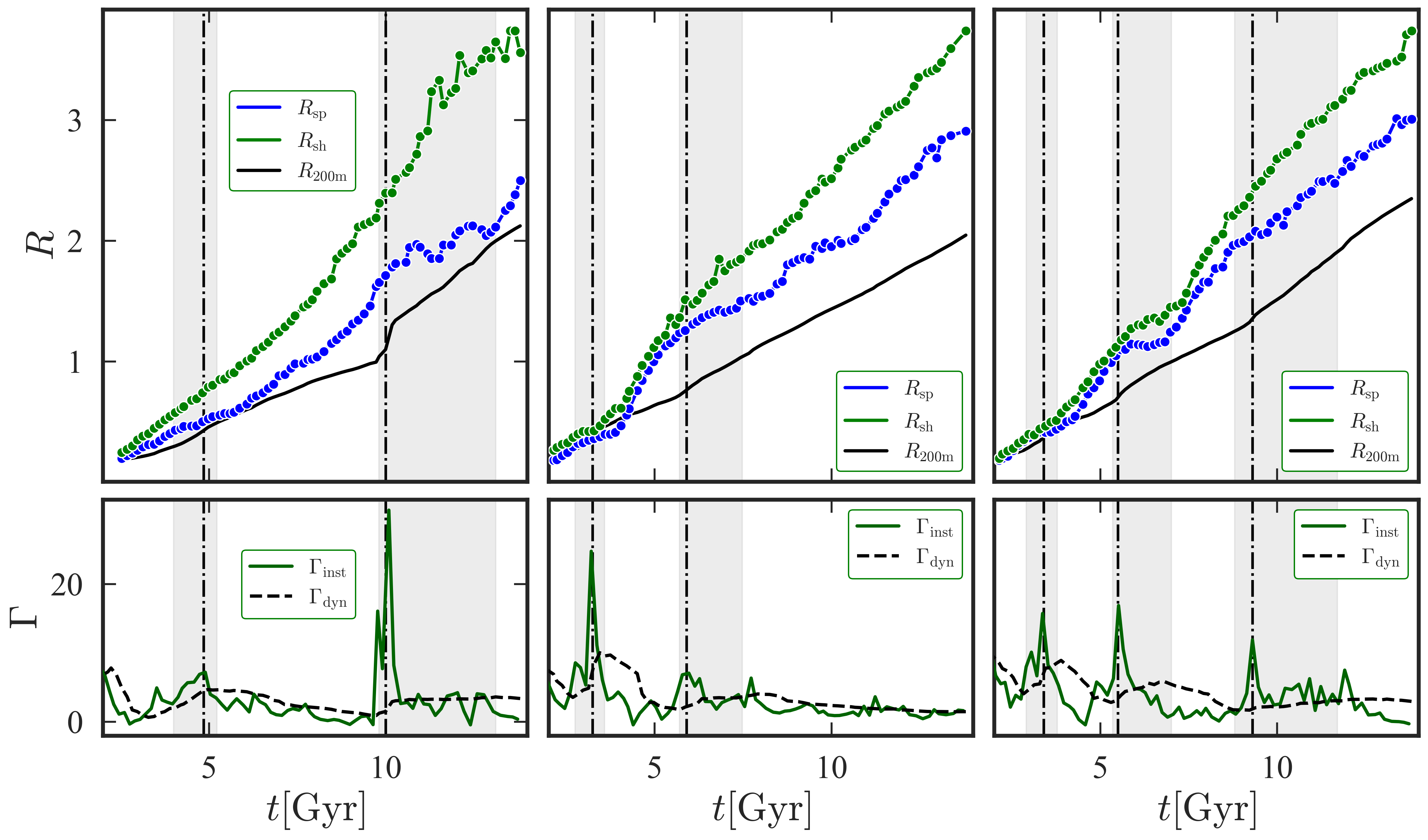}
    
    \caption{\textit{Top panel}: Evolution of $R_{\rm sh}$ , $R_{\rm sp}$ and $R_{\rm 200m}$ for clusters of masses \(M_{200m}=3.15 \times 10^{14}\, M_\odot\)(\emph{left}), \(5.6 \times 10^{14}\, M_\odot\) (\emph{middle}), and \(1.51 \times 10^{15}\, M_\odot\) (\emph{right}). \emph{Bottom panel}: The instantaneous and dynamical-time-averaged accretion rates are shown with green solid and black dashed lines, respectively, with dot dashed vertical lines representing merger events and shaded regions highlighting 1$t_{\rm dyn} $ from the merger. Both $R_{\rm sp}$ and $R_{\rm sh}$ contract after a merger, but $R_{\rm sh}$ settles at a higher cluster-centric radius. Overall, both $R_{\rm sp}$ and $R_{\rm sh}$ show a synchronous behaviour.}
    \label{fig:redshift_evolution}
\end{figure}

Figure~\ref{fig:redshift_evolution} illustrates the evolutionary history of three representative clusters. These halos have masses of $M_{\rm 200m}=3.15\times 10^{14} M_\odot$, $5.6\times 10^{14} M_\odot$, and $1.51\times 10^{15} M_\odot$ at $z=0$, with corresponding dynamical accretion rates ($\Gamma_{\rm dyn}$) of 3.37, 0.93, and 1.84. The top panel tracks the evolution of their splashback and shock radii, evaluated as the median of the angular distribution. The bottom panel shows the corresponding evolution of the instantaneous ($\Gamma_{\rm inst}$) and dynamical ($\Gamma_{\rm dyn}$) mass accretion rates as a function of redshift.

\begin{figure}[t]
    \centering
    \includegraphics[width=1\columnwidth]
    {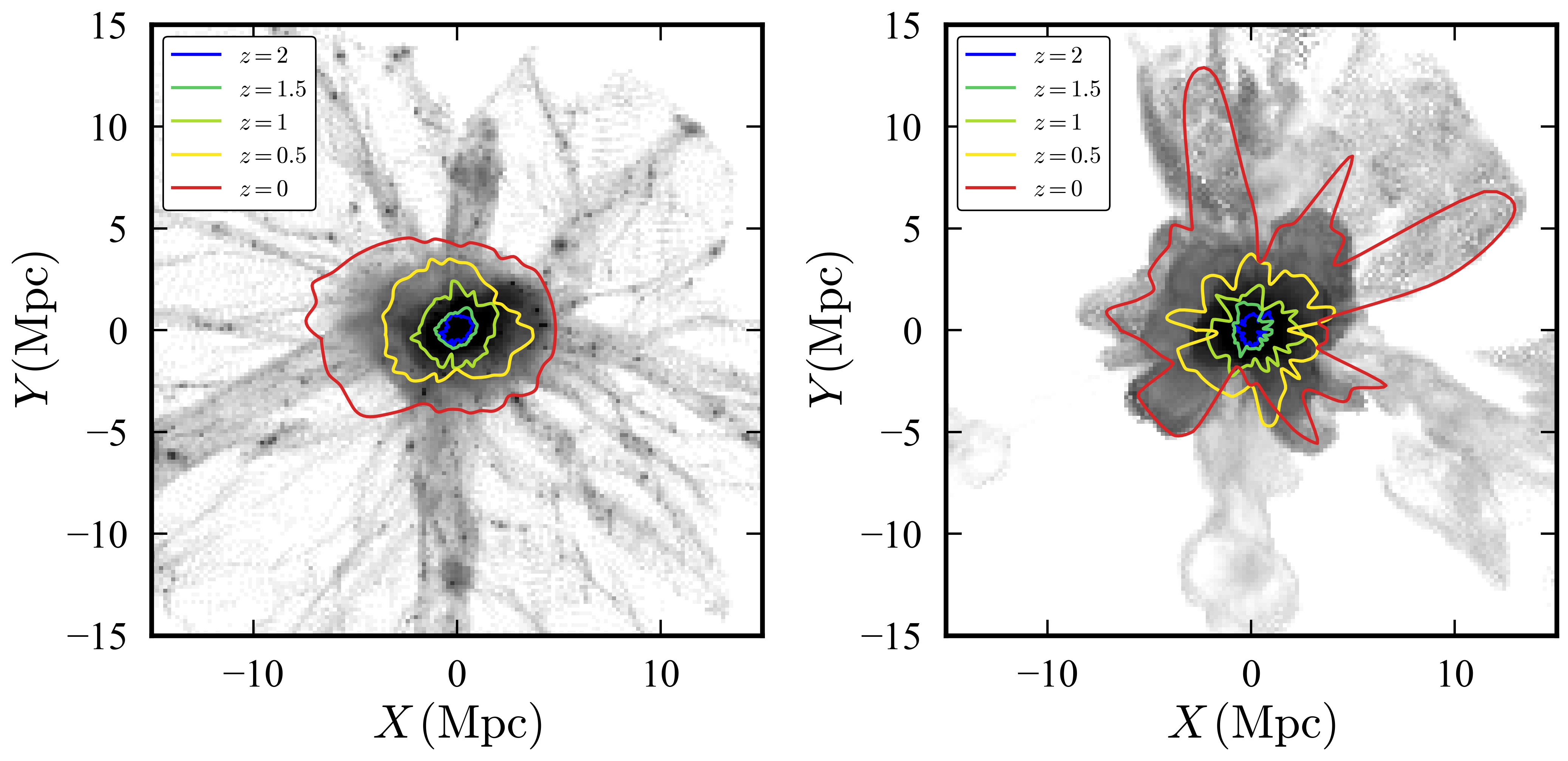}
    \caption{An example of a 2D projection of the splashback (\emph{left panel}) and shock (\emph{right panel}) surfaces as a function of redshift from $z=0-2$ for the most massive cluster \(M_{200m}=1.51 \times 10^{15}\, M_\odot\), overplotted on dark matter density (\emph{left}) and gas entropy (\emph{right}).}
    \label{fig:redshift_evolution_1}
\end{figure}

To provide a spatial visualisation of this evolution, Figure~\ref{fig:redshift_evolution_1} shows the projected splashback (left panel) and shock (right panel) surfaces overlaid on the projected dark matter density (left) and gas entropy (right) at several key redshifts.

The splashback radius ($R_{\rm sp}$) and the gas shock radius ($R_{\rm sh}$) exhibit strong oscillations driven by merger events that appear as peaks in the instantaneous accretion rate\footnote{Note that the actual values of the instantaneous merger rates are unphysical; they simply reflect that the halo finder accounts for the merger mass after specific time steps when the virial radii of the mergers cross-each other. However, the instantaneous accretion rate values are good visual guides of mergers.}. Initially, the two radii are not necessarily offset, as seen in the early evolution of one halo (Figure~\ref{fig:redshift_evolution}, third panel). During a merger, however, both radii contract on a timescale of $\sim$1~Gyr, consistent with the pericenter crossing time of an infalling halo. Notably, the splashback radius undergoes a more significant contraction than the shock boundary. This differential contraction introduces a persistent offset between the two radii. Following this phase, both boundaries expand, and each subsequent merger acts to further increase the separation between them.

Every merger perturbs the equilibrium state of dark matter and gas in the halo. The gas is in hydrostatic equilibrium, while the collisionless dark matter is thought to follow the Jeans equation in the fluid picture. Perturbations around equilibrium propagate differently in the two systems. Unlike dark matter, which relaxes through coherent processes such as Landau damping \cite{1967MNRAS.136..101L} and violent relaxation, the gas generates a true sound wave and can additionally be heated by shocks at merger boundaries through dissipative processes. This provides additional support against contraction, explaining the observed offset. This difference in response is likely the origin of the offset between the gas and dark matter boundaries.

The overall behaviour of the shock and splashback radii is consistent with that found in idealized simulations of merging clusters \cite{Zhang:2021wyi}, which also show the contraction and relaxation of these boundaries during mergers and possibly also during very rapid smooth accretion ($\Gamma\sim 3$) \cite{2020MNRAS.494.4539Z}. They attribute at least part of the expansion of the shock to shock acceleration driven by runaway merger shocks that originate near the centre and propagate beyond the virial shock. However, the synchronous behaviour of the shock and splashback radii—with different amplitudes—during both the contraction and relaxation phases also indicates breathing-mode like oscillations, which introduce subtle differences in the locations of the two boundaries. For instance, following a merger event (e.g., at $t \approx 6~\rm Gyr$ in the right panel of Figure~\ref{fig:redshift_evolution}), both boundaries contract, but the splashback radius contracts more significantly. During the subsequent relaxation, the shock expands to a larger fraction of $R_{\rm 200m}$, establishing the offset that persists to z=0. A detailed, quantitative study of the redshift evolution will be presented in a follow-up work. In particular, an extensive study of the redshift evolution of the full sample of clusters will give us a better understanding of the correlation between mergers and shock radius, additionally tracking the merger shocks along with the accretion shock, will help us disentangle the effects of acceleration due to runaway merger shocks and general perturbed oscillations of the boundaries.

\section{Conclusions}

We have presented measurements of the shock and splashback radii for galaxy clusters in the IllustrisTNG simulation suite. We use an entropy-based method to find the gas boundary and a novel method based on velocity dispersion to find the splashback boundary. Instead of a single spherical estimate, our algorithm finds the full directional gas and dark matter boundaries of the clusters. Our primary aim is to study the distribution of offsets between the shock and splashback boundaries of halos and their redshift evolution. Our main findings are summarised as follows:

\begin{itemize}

\item[(i)] The cluster edge shock is clearly imprinted in temperature, pressure, and entropy profiles, with entropy showing the sharpest peak at the shock location across all directions, including filaments.

\item[(ii)] At the boundary of the dark matter phase space, which separates the infall and multistreaming regions, the radial velocity dispersion shows a sharp transition along with the density. The dispersion transition occurs at the outermost radial distance of the phase space and is typically easier to measure in the low density region in the outskirts.

\item[(iii)] Shock distributions show greater scatter and larger median radii than splashback across all clusters. This offset, with a median ratio of $R_{\rm sh}/R_{\rm sp} \approx 1.3-2$, persists regardless of boundary definition, confirming it as a robust feature of cosmological hydrodynamical simulations. We also performed the analysis using the traditional definitions of splashback and shock radii and found an offset of $2.10^{+0.39}_{-0.26}$. We find, consistent with previous results, that the boundary of gas is offset from the dark matter boundary primarily along the void directions. Along the filamentary directions, the splashback surface and the gas shock boundary agree well with each other.

\item[(iv)]We explored various multidimensional dependencies of the shock feature and the splashback feature. We studied in detail the correlations of $R_{\rm sh}$ and $R_{\rm sp}$, normalised by $R_{\rm 200m}$ and their ratio ($\mathcal{R}\equiv R_{\rm sh}/R_{\rm sp}$) with halo properties such as halo mass ($\log M_{\rm 200m}$), accretion rate ($\Gamma$), and the steepness of the transition ($\mathcal{D}$) from the inner virialized region to the outer infall region. We find that $\log M_{\rm 200m}$ and  $\Gamma$ are the primary drivers for $R_{\rm sh}$ and $R_{\rm sp}$, with high accreting halos having lower $R_{\rm sh}/R_{\rm 200m}$ and $R_{\rm sp}/R_{\rm 200m}$.

\item[(v)]In units of $R_{\rm 200m}$, we find that $R_{\rm sh}$ decreases systematically with increasing halo mass. This correlation expressly violates the near scale-free behaviour that splashback exhibits, in the sense that $R_{\rm sp}/R_{\rm 200m}$ does not show a strong correlation with mass (barring a mild correlation with the accretion rate). However, more massive halos typically have a higher accretion rate. We binned the halos into 2D bins of mass and $\Gamma$ as shown in Figure~\ref{fig:2d_correlations}, and we found that indeed $R_{\rm sh}/R_{\rm 200m}$ decreases as the mass increases while keeping $\Gamma$ fixed.

\item[(vi)] To trace the origin of the Gas–DM offset, we analyse the high-time-resolution evolution of $R_{\rm sh}$, $R_{\rm sp}$ and $R_{\rm 200m}$ for three clusters from the TNG-Cluster sample spanning different masses and accretion rates. The offset emerges early and remains small. Mergers amplify it but are not the sole driver: after a merger, both boundaries decrease and then rise nearly in sync, but the shock settles at a larger radius, widening the offset.

\item[(vii)] With regard to observability of the feature we find that taking an angular average over different directions tends to wash out the outer shock from the pressure profile for typical cluster mass halos. The integrated pressure is the primary observable in SZ surveys.  This smearing happens because the angular average is density-weighted and the gas density by itself does not show a sharp feature.  Measuring profiles along void and filamentary directions separately, or finding angular medians in data, will allow us to extract information about the existence of the outer accretion shock.

\end{itemize}

Our work demonstrates that, in hierarchically forming clusters, a persistent offset develops between the gas shock and the dark matter splashback radius. Studying this phenomenon is crucial for understanding the halo-gas connection, which is necessary for robustly modelling the cluster outskirts. This offset is not static; it is introduced and amplified by merger events, and its measurement encodes information about the dynamical state of the cluster. Probing this effect with multi-wavelength surveys provides a powerful new window into the 1- to 2-halo transition regime, where the cluster environment connects to the cosmic web. While detecting these low-density outer regions is challenging, initial breakthroughs are being made with stacked measurements of the thermal SZ effect and X-ray. However, the full diagnostic potential of this effect will be unlocked by a suite of next-generation probes. Exploring this offset with the kinematic and relativistic SZ effects and FRBs, as well as extending the analysis to lower-mass clusters, is a compelling avenue for future research.

\acknowledgments
We thank the IllustrisTNG project for making the TNG-Cluster and the flagship TNG simulation suites publicly available. We thank Arka Banerjee, Aseem Paranjpe, Ravi Sheth and Congyao Zhang for discussions. We thank Jiaxin Han, Erwin Lau, Phil Mansfield, Dylan Nelson, Annalisa Pilepich, Michele Pizzardo and Mingtao Yang for comments on the draft. We received support from the PARAM Brahma supercomputing facility at IISER Pune, part of the National Super Computing Mission under the Government of India. SA was supported under ANRF grant SRG/2023/001563. 

\appendix

\section{Resolution Study}
\label{appendix:resolution_study}

To test the numerical convergence of our shock and splashback detection algorithms, we analyse the 100 most massive cluster-scale halos ($M > 10^{14} M_\odot$) from three resolution levels of the IllustrisTNG simulation. We use TNG300-1 (high-resolution), TNG300-2 (8$\times$ lower mass resolution), and TNG300-3 (64$\times$ lower mass resolution), which correspond to dark matter particle masses of $5.9 \times 10^{7}, M_\odot$, $4.7 \times 10^{8}, M_\odot$, and $3.8 \times 10^{9}, M_\odot$, respectively.

\begin{figure}[h] 
   
    \hspace*{-0.02\linewidth}  
    \includegraphics[width=1.02\columnwidth]{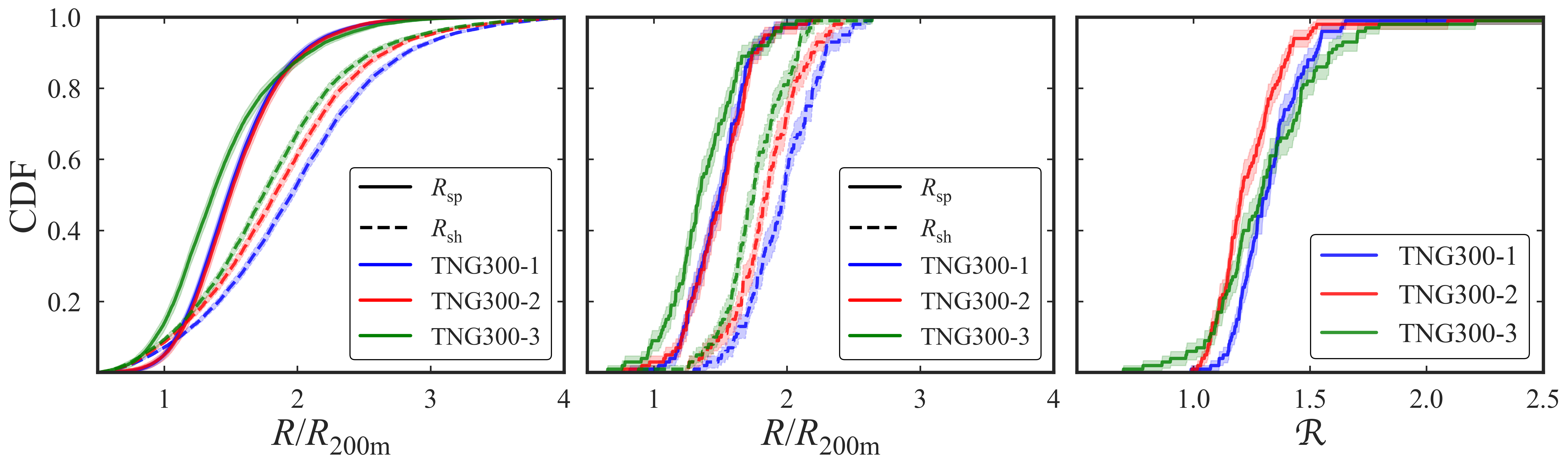} 
    \caption{Resolution dependence of splashback (solid lines) and shock (dashed lines) boundary statistics for the 100 most massive clusters ($M_{200\text{m}} > 10^{14} M_\odot$) in the TNG300 simulation suite, comparing TNG300-1 (blue), TNG300-2 (red), and TNG300-3 (green).
    \emph{Left panel:} Cumulative Distribution Function (CDF) of all directional radii combined from the entire cluster sample, normalized by $R_{\rm 200m}$.
    \emph{Middle panel:} CDF of the median radius for each individual cluster.
    \emph{Right panel:} CDF of the ratio $\mathcal{R}\equiv R_{\rm sh}/R_{\rm sp}$ computed using the median radii for each cluster.
    }
    \label{fig:jackknife}
\end{figure}

Figure~\ref{fig:jackknife} presents our resolution convergence test, highlighting that the splashback boundary is more numerically stable than the shock boundary. The left panel shows the cumulative distribution functions (CDFs) of the boundary radii measured at different angles; here, the splashback boundary demonstrates excellent stability between the TNG300-1 and TNG300-2 runs, while the shock detection exhibits stronger resolution dependence. To quantify this, the middle panel shows the CDFs of the median radii derived from these angular distributions, while the right panel displays the CDFs of their ratios. Shaded regions in all panels indicate the uncertainties estimated via the jackknife method.

\begin{figure}[t] 
    \centering
   
    \includegraphics[width=0.77\columnwidth]{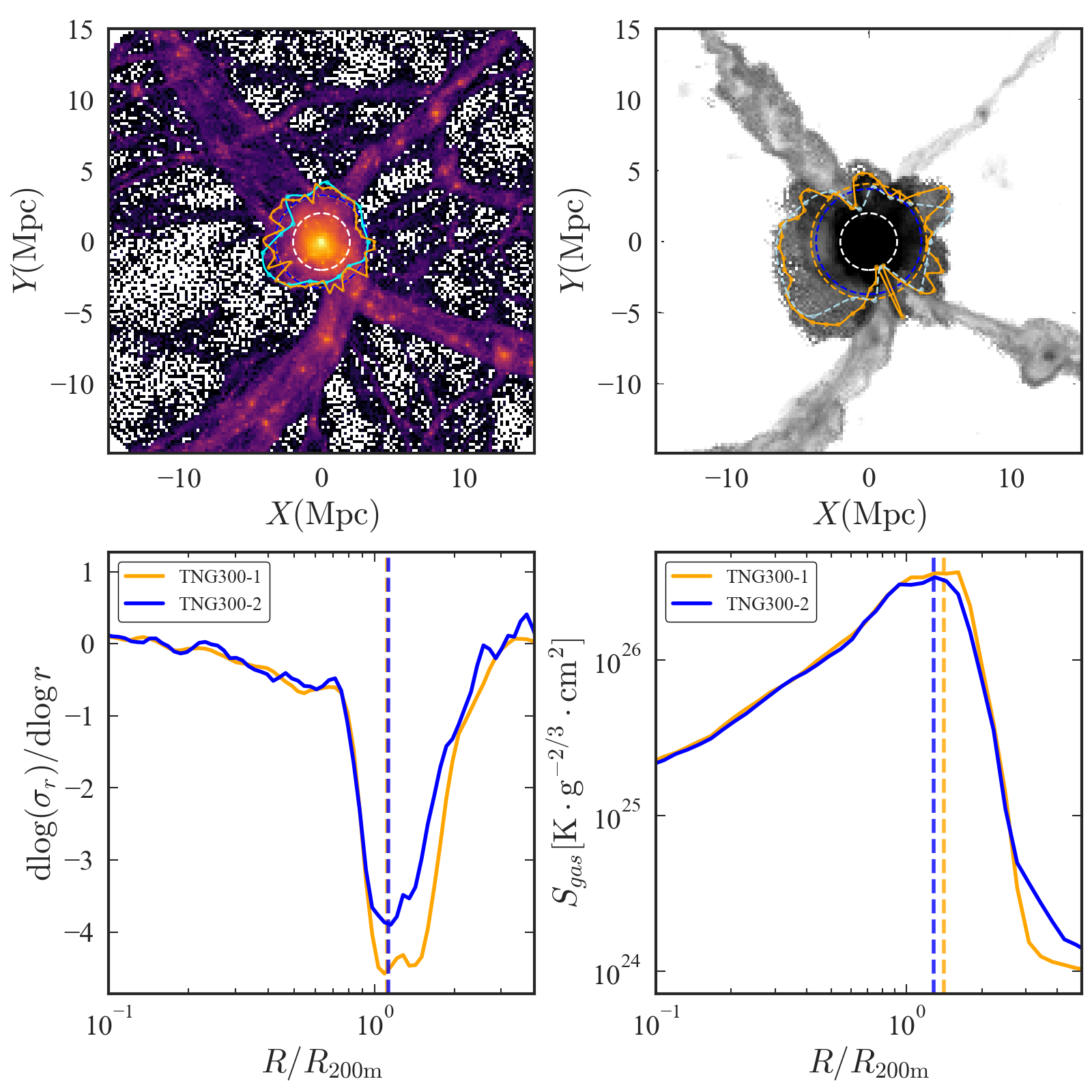} 
    \caption{Resolution convergence test for the shock and splashback boundaries. This case study shows a single cluster ($M_{\rm 200m} = 2.62 \times 10^{14} M_\odot$) from TNG300-1 (high-resolution, orange) and TNG300-2 (medium-resolution, light blue). 
    \emph{Top Panels:} 2D projections of the dark matter density (left) and gas entropy (right), overlaid with the corresponding splashback and shock boundaries. The white dashed circle marks $R_{\rm 200m}$. 
    \emph{Bottom Panels:} The 1D radial profiles used for boundary detection. The splashback radius ($R_{\rm sp}$) is identified from the minimum of the velocity dispersion slope (left), while the shock radius ($R_{\rm sh}$) is identified from the maximum of the entropy profile (right). Dashed vertical lines indicate the measured radii.
    }
    \label{fig:halo_12}
\end{figure}

Figure~\ref{fig:halo_12} provides a detailed case study of a representative cluster ($M \approx 2.62 \times 10^{14}, M_\odot$) to visually illustrate the impact of resolution by comparing its properties in TNG300-1 and TNG300-2. The upper panels display 0.7 Mpc thick 2D slices of the halo. In the top-left, the splashback boundaries for TNG300-1 (orange) and TNG300-2 (light blue) are superimposed on the dark matter density map. The top-right similarly overlays the shock boundaries on the gas entropy map using the corresponding colour coding. The lower panels present the 1D radial profiles from which these boundaries are derived: the logarithmic derivative of the radial velocity dispersion (bottom-left) to find the splashback radius and the median entropy profile (bottom-right) to find the shock radius. Vertical lines mark the locations of these detected radii, and the cluster's $R_{\rm 200m}$ is indicated by a white dashed circle in all panels for reference.

\begin{figure}[t] 
   \includegraphics[width=1\columnwidth]{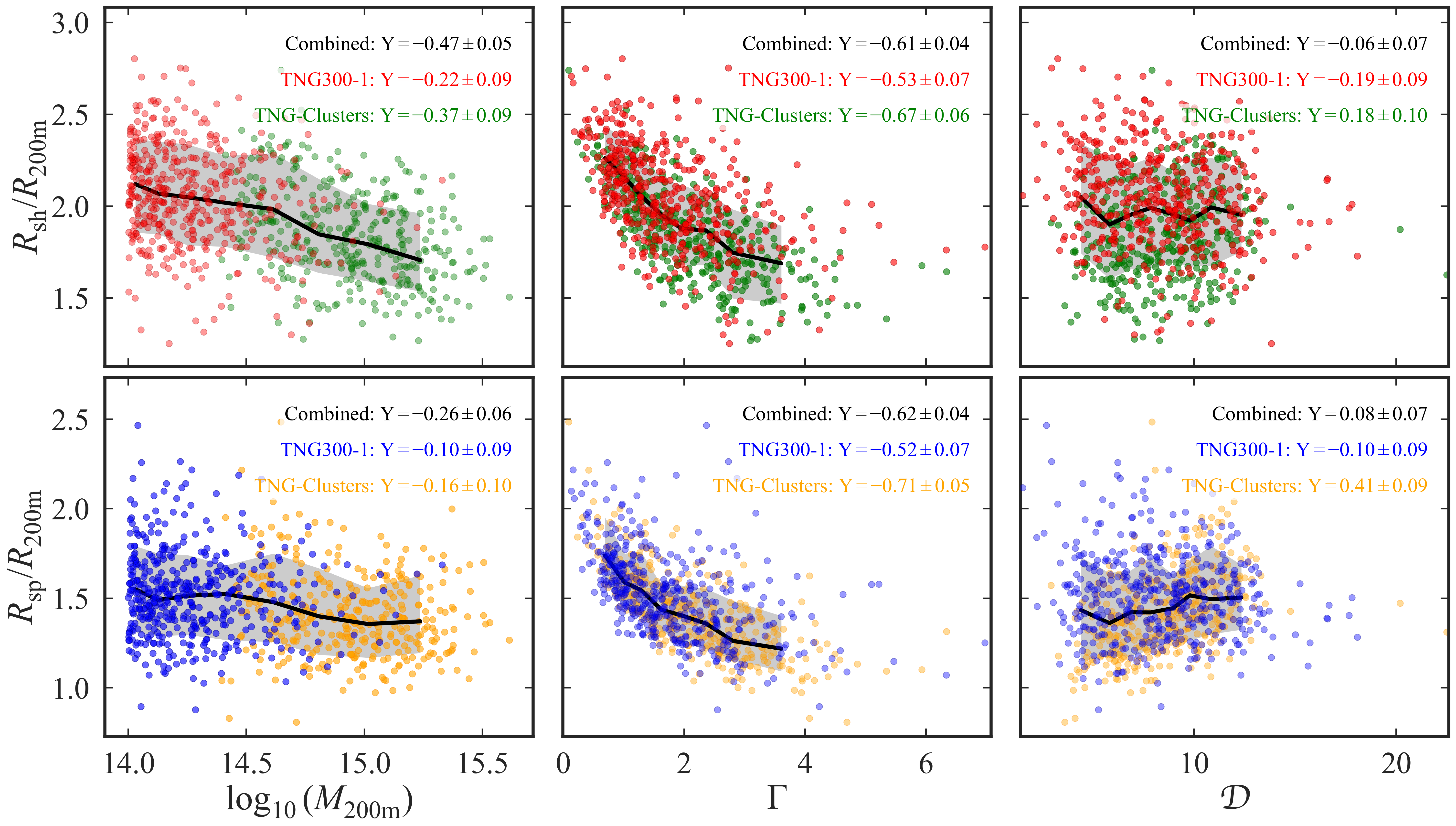} 
    \caption{The relationship between halo properties and the shock and splashback radii. The top row shows the shock radius ($R_{\rm sh}$) and the bottom row shows the splashback radius ($R_{\rm sp}$), both normalised by $R_{\rm 200m}$. From left to right, each radius is plotted against halo mass ($M_{200m}$), mass accretion rate ($\Gamma$), and dispersion depth ($\mathcal{D}$). Data is from a sample of galaxy clusters in the TNG300-1 and TNG-Cluster simulations. This comparison highlights how each boundary traces different aspects of halo growth.
    }
    \label{fig:clust_vs_300}
\end{figure}

We investigate the relationship between halo boundaries and assembly history using clusters from the TNG300-1 and TNG-Cluster simulations. Figure~\ref{fig:clust_vs_300} shows the normalised shock ($R_{\rm sh}/R_{\rm 200m}$) and splashback ($R_{\rm sp}/R_{\rm 200m}$) radii as a function of halo mass ($M_{200m}$), mass accretion rate ($\Gamma$), and dispersion depth ($\mathcal{D}$). Consistent with theoretical expectations, both radii show a strong anticorrelation with the accretion rate, decreasing in halos that are assembling mass more rapidly. However, the two boundaries exhibit distinct dependencies on halo mass. The splashback radius is largely scale-free, whereas the shock radius shows a clear negative trend with mass, which becomes stronger in the more massive TNG-Cluster sample. Furthermore, a new scaling relation emerges for the splashback radius in these massive clusters: $R_{\rm sp}$ correlates strongly with dispersion depth, a proxy for the dispersion profile's sharpness. This correlation is not observed in the lower-mass TNG300-1 halos, indicating that internal halo structure becomes a more significant factor in setting the splashback boundary at the highest mass scales.

\section{Average Gas Pressure Profiles along Voids and Filaments}
\label{appendix:environment}

\begin{figure}[H] 
    \centering
    \includegraphics[width=1\columnwidth]{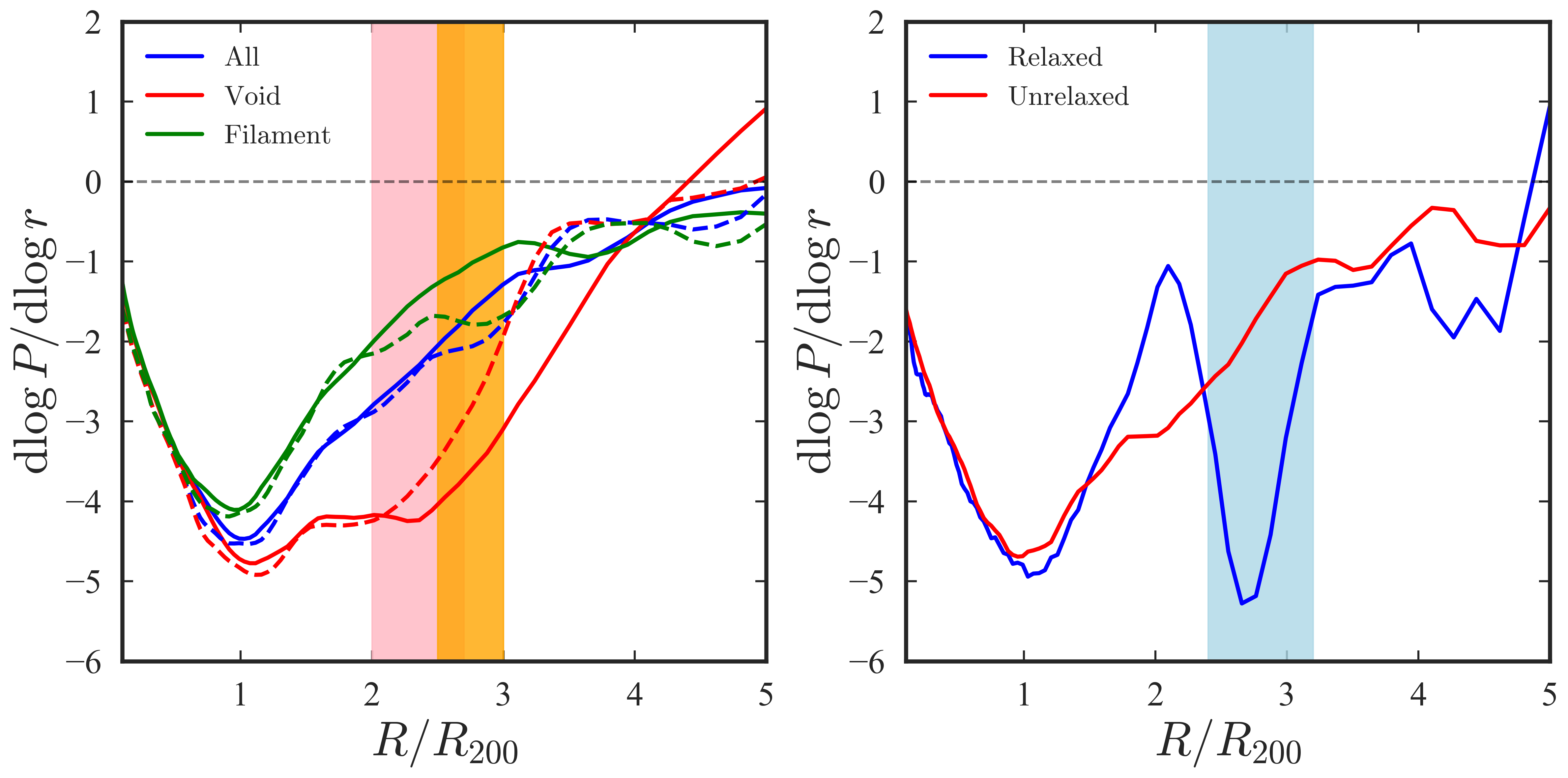} 
    \caption{{\emph{Left:} Logarithmic derivative of stacked  averaged pressure profiles of gas for clusters from TNG300-1 (solid) and TNG-Cluster (dashed) lines split along voids and filaments. 
    \emph{Right:} Logarithmic derivative of stacked averaged pressure profiles for relaxed ($\Gamma_{\mathrm{dyn}} < 1$) and unrelaxed ($\Gamma_{\mathrm{dyn}} > 1$) clusters with $M > 10^{15} M_\odot$ from TNG-Cluster. The shaded regions indicate the shock region between the minima of the logarithmic derivative of entropy and its maxima for TNG300-1 and TNG-Cluster.}}
    \label{fig:void}
\end{figure}

To investigate the dependence of shock detectability of clusters on their large scale environment, we examine the logarithmic derivative of stacked averaged gas pressure profiles, split according to filamentary or void directions. Figure~\ref{fig:void} (left) shows these derivatives for clusters from TNG300-1 (solid lines) and TNG-Cluster (dashed lines), with all directions combined; void and filament directions are shown in blue, red, and green, respectively. The pink and orange shaded regions correspond to the shock region between the minimum of the logarithmic slope and the profile maximum of entropy for TNG-Cluster and TNG300, respectively.\\
To assess how the dynamical state of clusters influences the detectability of accretion shocks, we further split the TNG-Cluster sample by mass and accretion rate. Specifically, we select clusters with $M > 10^{15}  M_\odot$ and classify them as relaxed ($\Gamma_{\mathrm{dyn}} < 1$) or unrelaxed ($\Gamma_{\mathrm{dyn}} > 1$) based on their accretion rate. The right panel of Figure~\ref{fig:void} presents the derivative of the stacked angular-averaged pressure profiles for these two populations. Relaxed clusters exhibit a well-defined minimum in the pressure derivative, corresponding to a sharp accretion shock, whereas unrelaxed clusters show a much weaker feature. A similar trend was observed in relaxed clusters from the ThreeHundred simulations by \citep{Baxter:2021tjr}.\\ 
The steepest gradient of the profile, commonly associated with the shock radius, is substantially sharper in void directions than along filaments. Accretion along filaments tends to smear out the sharp edge expected from a coherent accretion shock, making it more difficult to detect shocks in angular-averaged profiles. In contrast, void directions appear to host more quiescent accretion, resulting in a cleaner dynamical boundary. These massive clusters often dominate their surroundings, and when accretion from filaments is subdued, they develop a prominent outer shock that may be detectable in future SZ surveys.

\bibliography{shock}
\bibliographystyle{JHEP}

\end{document}